\documentclass[aps,prb,reprint]{revtex4-1}

\usepackage{graphicx}
\usepackage{amsmath}
\usepackage{amssymb}
\usepackage{multirow}

\newcommand{\sub}[1]{_{\ensuremath{\textrm{#1}}}} 
\newcommand{\super}[1]{\ensuremath{^{\textrm{#1}}}} 
\newcommand{\plustimes}{\ensuremath{+\hspace{-7.7pt}\times}} 

\begin{document}

\title{Ideal regularization of the Coulomb singularity in exact exchange by
Wigner-Seitz truncated interactions: towards chemical accuracy in non-trivial systems}

\author{Ravishankar Sundararaman}
\author{T. A. Arias}
\affiliation{Department of Physics, Cornell University, Ithaca, New York 14853, USA}
\date{\today}

\begin{abstract}
Hybrid density functionals show great promise for chemically-accurate first principles calculations,
but their high computational cost limits their application in non-trivial studies,
such as exploration of reaction pathways of adsorbents on periodic surfaces.
One factor responsible for their increased cost is the dense Brillouin-zone sampling
necessary to accurately resolve an integrable singularity in the exact exchange energy.
We analyze this singularity within an intuitive formalism based on Wannier-function
localization and analytically prove Wigner-Seitz truncation to be the ideal method
for regularizing the Coulomb potential in the exchange kernel.
We show that this method is limited only by Brillouin-zone discretization errors in the
Kohn-Sham orbitals, and hence converges the exchange energy exponentially with the number of
$k$-points used to sample the Brillouin zone for all but zero-temperature metallic systems.
To facilitate the implementation of this method, we develop a general construction for the
plane-wave Coulomb kernel truncated on the Wigner-Seitz cell in one, two or three lattice directions.
We compare several regularization methods for the exchange kernel in a variety of
real systems including low-symmetry crystals and low-dimensional materials.
We find that our Wigner-Seitz truncation systematically yields the best $k$-point convergence
for the exchange energy of all these systems and delivers an accuracy to hybrid functionals
comparable to semi-local and screened-exchange functionals at identical $k$-point sets.
\end{abstract}
\maketitle

\section{Introduction} \label{sec:Introduction}

Density-functional theory\cite{HK-DFT,KS-DFT} forms the basis for \emph{ab initio}
theoretical studies of the ground-state electronic structure of materials, and
serves as the starting point for many-body perturbation theories such as
GW\cite{GW} and Bethe-Salpeter equation\cite{BSE} (BSE) calculations.
Standard semi-local approximations to exchange and correlation in density-functional theory,
such as the local-density and generalized-gradient approximations, are remarkably accurate
for a variety of properties such as lattice constants, equilibrium geometries and elastic moduli,
but are not sufficiently accurate for the energetics and kinetics of chemical reactions.\cite{DFTinaccuracy}

Hybrid density functionals, which replace a fraction of the approximate semi-local exchange energy
with the exact non-local Fock exchange energy, improve upon the accuracy of semi-local functionals
and have been widely applied for first-principles thermo-chemistry.\cite{B3LYP}
Variants of these functionals\cite{PBE0} enable calculations for solids and surfaces
with accuracy sufficient for predicting atomic-scale processes at room temperature.
However, hybrid functionals require a greater number of $k$-points than semi-local functionals
for comparable accuracy in Brillouin-zone discretization for periodic systems.
This increases their already high computational cost and limits their applicability, so that
practical studies of surface reactions and phenomena such as catalysis remain tantalizingly out of reach.

The need for finer $k$-point sampling in hybrid functionals stems from the Brillouin-zone
integrals over the singular Coulomb kernel in the exact exchange energy of periodic systems.
Similar operators also appear in GW and BSE calculations, and so successful methods
to address this issue have implications for excited state methods as well.

The discretization error in singular reciprocal space integrals critically
depends on the technique used to handle the singular contributions.
The standard auxiliary-function approach\cite{AuxFunc-GygiBaldereschi,AuxFunc-Wenzien,AuxFunc-Carrier}
replaces the divergent terms with an average around the singularity computed
using an auxiliary function with the same singularity as the Coulomb kernel.
However, such methods that replace only the divergent terms lead to polynomial convergence
($N_k^{-1}$) with the number of $k$-points $N_k$,\cite{AuxFunc-Carrier} compared to the
standard exponential convergence ($\exp(-a N_k^{1/3})$) in the case of semi-local functionals.

Some hybrid functionals\cite{HSE03} achieve $k$-point convergence comparable to fully semi-local
functionals by replacing the exact exchange energy with a screened exchange energy computed using
a short-ranged kernel $\textrm{erfc}(\omega r)/r$ that is not singular at long wave-vectors.
The remaining long-ranged part $\textrm{erf}(\omega r)/r$ is treated using a semi-local (generalized-gradient) approximation.
The predictions of these functionals depend on the screening parameter $\omega$ due to this additional approximation, and are
less accurate than those of exact-exchange hybrids for some properties such as elastic constants of periodic systems.\cite{HSE12}
A method with comparable computational cost but with the exact $1/r$ kernel would therefore be highly valuable.

One approach that shows promising results for the exact exchange energy avoids the
singularity by imposing a real-space cutoff on the Coulomb kernel with a length-scale
dependent on the $k$-point mesh.\cite{SphericalTruncation,QMC-MPC,ModelAnalysisEXX}
The spherical truncation employed in this approach, however, works well only for high-symmetry crystals,
whose Wigner-Seitz cells more or less resemble spheres. The valuable gains afforded by such an approach
in these special cases indicates the need for a detailed understanding of why truncated potentials work
for calculating the exchange energy of periodic systems so as to point the way to a general method applicable to all systems.

Following this program, Section~\ref{sec:RealSpaceAnalysis} analyzes the singularity in the exact
exchange energy of periodic systems using a formalism based on Wannier-function localization.
This allows us to prove analytically that Wigner-Seitz truncation of the Coulomb potential
is the ideal regularization method with accuracy limited only by Brillouin-zone
discretization errors in the Kohn-Sham orbitals themselves; the appendix develops
a set of techniques necessary to truncate the Coulomb potential on Wigner-Seitz cells.
Section~\ref{sec:PartiallyPeriodicExchange} then generalizes regularization methods
for the exchange energy in three-dimensional systems to slab-like (two-dimensional)
and wire-like (one-dimensional) geometries using partially-truncated potentials.
Finally, Section~\ref{sec:Results} compares truncated-potential and auxiliary-function approaches
for a variety of real materials with high- and low-symmetry crystal systems, electronic structure
ranging from insulating to metallic, and dimensionality ranging from three to one.

The results indicate that employing Wigner-Seitz truncated Coulomb kernels is systematically
more accurate than other methods for dealing with the singularity in Fock exchange.
Moreover, at equivalent Brillouin-zone sampling, the accuracy of hybrid functionals
which include exact exchange computed using the Wigner-Seitz truncated method rivals
that of functionals that only include screened exchange, and even that of {\it semi-local}
functionals which include {\it no} non-local exchange contributions {\it whatsoever}.

\section{Exchange in periodic systems} \label{sec:ExchangeFormalism}

The exact Fock exchange energy of a finite system with Kohn-Sham orbitals
$\psi_{i\sigma}(\vec{r})$ and corresponding occupation numbers $f_{i\sigma}$ is
given by the non-singular expression,
\begin{multline}
E_X = \frac{-1}{2} \sum_{i,j,\sigma} f_{i\sigma} f_{j\sigma} \int d\vec{r} \int d\vec{r}' \\
	\times \frac{\psi_{i\sigma}(\vec{r})\psi_{j\sigma}^{\ast}(\vec{r})
		\psi_{i\sigma}^{\ast}(\vec{r}')\psi_{j\sigma}(\vec{r}')}{|\vec{r}-\vec{r}'|}.
\label{eqn:ExchangeFinite}
\end{multline}
Here, we work with atomic units $e^2/(4\pi\epsilon_0)=\hbar^2/m_e=1$,
so that the unit of distance is the Bohr ($a_0$) and the unit of energy is the Hartree ($E_h$).
The exchange energy is always a sum of independent contributions for each spin channel;
the rest of this section omits the spin index $\sigma$ and the implicit sum over $\sigma$ for clarity.

In a periodic system, the Kohn-Sham orbitals take the Bloch form
$\psi_i^{\vec{k}}(\vec{r}) = e^{i\vec{k}\cdot\vec{r}} u_i^{\vec{k}}(\vec{r})$,
where $u_i^{\vec{k}}(\vec{r})$ are periodic functions normalized on the unit cell
of volume $\Omega$ and labeled by band index $i$ as well as a wave-vector
$\vec{k}$ in the Brillouin zone.
The exchange energy (\ref{eqn:ExchangeFinite}) of the periodic system
per unit cell (per spin) is
\begin{multline}
E_X = \frac{-1}{2} \sum_{i,j}
	\int\sub{BZ}\frac{\Omega d\vec{k}}{(2\pi)^3} \int\sub{BZ}\frac{\Omega d\vec{k}'}{(2\pi)^3}
	f_i^{\vec{k}} f_j^{\vec{k}'} \int_\Omega d\vec{r} \int d\vec{r}' \\
		\times \frac{\psi_i^{\vec{k}}(\vec{r})\psi_j^{\vec{k}'\ast}(\vec{r})
		\psi_i^{\vec{k}\ast}(\vec{r}')\psi_j^{\vec{k}'}(\vec{r}')}{|\vec{r}-\vec{r}'|},
\label{eqn:ExchangePeriodic}
\end{multline}
where $\int\sub{BZ}$ denotes integration over the Brillouin zone.
The conventional treatment of this energy begins
with a plane-wave expansion of the product densities
\begin{equation}
\psi_i^{\vec{k}\ast}(\vec{r}) \psi_j^{\vec{k}'}(\vec{r})
	\equiv  \rho^{\vec{k}\vec{k}'}_{ij}(\vec{r})
	= e^{i(\vec{k}'-\vec{k})\cdot\vec{r}} \sum_{\vec{G}}
		e^{i\vec{G}\cdot\vec{r}} \tilde{\rho}^{\vec{k}\vec{k}'}_{ij\vec{G}},
\end{equation}
where $\tilde{\rho}^{\vec{k}\vec{k}'}_{ij\vec{G}}$ are the Fourier
components of those densities at reciprocal lattice vectors $\vec{G}$.
This treatment then rewrites (\ref{eqn:ExchangePeriodic}) in Fourier space as
\begin{equation}
E_X = \frac{-\Omega}{2}
	\int\sub{BZ}\frac{\Omega d\vec{k}}{(2\pi)^3} \int\sub{BZ}\frac{\Omega d\vec{k}'}{(2\pi)^3}
	\sum_{i,j,\vec{G}}
		f_i^{\vec{k}}f_j^{\vec{k}'}
		|\tilde{\rho}^{\vec{k}\vec{k}'}_{ij\vec{G}}|^2
		\tilde{K}_{\vec{G}+\vec{k}'-\vec{k}},
\label{eqn:ExchangePWinfinite}
\end{equation}
where the periodic Coulomb kernel $\tilde{K}_{\vec{q}} \equiv 4\pi/q^2$,
so that for each $\vec{k}'$ at $\vec{G}=0$, the integral over
$\vec{q}=\vec{k}-\vec{k}'$ is singular at $\vec{q}=0$.
This singularity is integrable since near $q=0$, the integral
$\sim \int 4\pi q^2 dq \frac{1}{q^2}$.

The above approach, however, is problematic for any practical calculation where
Brillouin-zone integrals are approximated using a finite quadrature, that is, as
a weighted sum over a set of `$k$-points'. In this paper, we restrict our attention
to the commonly employed Gauss-Fourier quadratures, which correspond to
uniform $k$-point meshes such as the Monkhorst-Pack grid.\cite{MonkhorstPack}
The exchange energy computed in practice is therefore
\begin{equation}
E_X = \frac{-\Omega}{2N_k^2} \sum_{\vec{k},\vec{k}',i,j,\vec{G}}
		f_i^{\vec{k}}f_j^{\vec{k}'}
		|\tilde{\rho}^{\vec{k}\vec{k}'}_{ij\vec{G}}|^2
		\tilde{K}_{\vec{G}+\vec{k}'-\vec{k}}.
\label{eqn:ExchangePW}
\end{equation}
where $N_k$ is the total number of $k$-points used for Brillouin zone sampling.
In principle, the exchange energy would converge with increasing density
of $k$-points even if the singular terms are dropped, or equivalently,
the Coulomb kernel is regularized with $\tilde{K}_{q=0}=0$, as usual.
However, that results in an $\mathcal{O}(\delta q)$ error, where
 $\delta q$ is the typical distance between neighboring $k$-points,
which leads to an impractically slow $N_k^{-1/3}$ convergence.

Auxiliary-function methods\cite{AuxFunc-GygiBaldereschi} address this poor
Brillouin zone convergence of the Fock exchange energy by choosing a value for the 
$G=0$, $\vec{k}=\vec{k}'$ term in (\ref{eqn:ExchangePW}) that captures the average
contribution of $4\pi/|\vec{k}'-\vec{k}|^2$ in the neighborhood of $\vec{k}=\vec{k}'$.
These methods correct for the finite quadrature error by setting this term to the difference between the
exact integral and the discrete $k$-point sum over the Brillouin zone of a function $f(q)$ that matches
the periodicity and the $4\pi/q^2$ singularity of the integrand in (\ref{eqn:ExchangePWinfinite}).
For a uniform $k$-point mesh, this amounts to replacing the Coulomb kernel
$\tilde{K}_{\vec{q}}$ in (\ref{eqn:ExchangePW}) with
\begin{equation}
\tilde{K}_{\vec{q}}\super{aux} = 
\begin{cases}
	4\pi/q^2,& \vec{q}\neq 0 \\
	N_k \int\sub{BZ} \frac{\Omega d\vec{k}}{(2\pi)^3} f(\vec{k})
	- \sum_{\delta\vec{k}} f(\delta\vec{k})
		,& \vec{q}=0,
\end{cases}
\end{equation}
where the discrete sum runs over $\delta\vec{k}$
in the $k$-point difference mesh excluding the $\Gamma$ point.\footnote{
For a uniform $k$-point mesh, the difference mesh is uniform
and $\Gamma$-centered even if the original mesh is off-$\Gamma$.}

The original method of Gygi and Baldereschi presented such a function for the
face-centered cubic lattice which could be integrated analytically.
Wenzien and coworkers\cite{AuxFunc-Wenzien} constructed similar functions for a few
other lattice systems and tabulated the corresponding $q=0$ corrections numerically.
Carrier and coworkers\cite{AuxFunc-Carrier} constructed a general function
that works for all Bravais lattices and prescribed a general scheme for computing
the Brillouin zone integral. Below, when making comparisons to our truncated potential method,
we employ this last variant of the the auxiliary-function method due to its generality,
and refer the reader interested in further details of auxiliary-function methods to Ref.~\citenum{AuxFunc-Carrier}.

With the correctly chosen $G=0$ term for $\vec{k}=\vec{k}'$, the auxiliary function
methods achieve $N_k^{-1}$ convergence\cite{AuxFunc-Carrier} in the exchange energy.
Duchemin and Gygi\cite{AuxFunc-DucheminGygi} have generalized the method
to achieve $N_k^{-2}$ convergence by introducing `curvature corrections' which,
in our notation above, amount to correcting the Coulomb kernel at $q=0$ as well as
$\vec{q}$ that correspond to nearest neighbor displacements in the $k$-point mesh.
For many systems, this method yields reasonable accuracy with modest $k$-point meshes;
however the asymptotic polynomial convergence of the exchange energy is still slower
than the exponential convergence of the total energy of semi-local density functionals.
We next describe a method for the exchange energy that achieves this exponential convergence.

\subsection{Real-space analysis of asymptotic convergence} \label{sec:RealSpaceAnalysis}

An alternate scheme to improve the Brillouin zone convergence of exact exchange
imposes a real-space cutoff on the Coulomb kernel in (\ref{eqn:ExchangePW}).
This scheme been shown to work reasonably well for high-symmetry crystals,\cite{SphericalTruncation}
but the reasons for its success remain somewhat mysterious. Two possible explanations have been offered.
The first is that the method satisfies the normalization constraints of the
exchange hole by appropriately truncating the Coulomb potential.\cite{QMC-MPC}
The second is that the effective distinguishability of electrons amongst different $k$-point
sampled supercells requires suppression of the exchange interaction between supercells.\cite{SphericalTruncation}
These explanations do not elucidate the underlying reason for an infinite-range interaction to be best
numerically approximated by a finite-range one nor specify what form that finite-ranged interaction should take,
and they do not lend themselves to an analysis of the accuracy or convergence properties of such an approximation.
To provide such an explanation and to identify the ideal form of the truncation, we now analyze
the Fock exchange interaction computed on finite $k$-point meshes in real space, and show that
the need for truncating the Coulomb potential arises both naturally and in a particular form.

We start by rearranging the exchange energy of the periodic system
(\ref{eqn:ExchangePeriodic}) as
\begin{equation}
E_X = \frac{-1}{2} \sum_{i,j,\vec{R}} \int d\vec{r}  \int d\vec{r}'
	\frac{ \bar{\rho}^{\vec{R}\ast}_{ij}(\vec{r}) \bar{\rho}^{\vec{R}}_{ij}(\vec{r}') }{ |\vec{r}-\vec{r}'| },
\label{eqn:ExchangeWannier}
\end{equation}
a sum of Coulomb self-energies of the pair densities $\bar{\rho}^{\vec{R}}_{ij}(\vec{r})
= \bar{w}_i^{\vec{0}\ast}(\vec{r}) \bar{w}_j^{\vec{R}}(\vec{r})$ of the Wannier-like functions
\begin{equation}
\bar{w}_i^{\vec{R}}(\vec{r}) = \int\sub{BZ} \frac{\Omega d\vec{k}}{(2\pi)^3}
	e^{-i\vec{k}\cdot\vec{R}} \psi_i^{\vec{k}}(\vec{r}) \sqrt{f_i^{\vec{k}}}.
\label{eqn:WannierLike}
\end{equation}
Indeed, order-$N$ calculations of the exchange energy\cite{WannierEXX} using
maximally-localized Wannier functions\cite{MLWF} employ similar transformations.
In contrast, we use (\ref{eqn:ExchangeWannier}), which is exactly equivalent to
(\ref{eqn:ExchangePeriodic}), only as a tool to analyze standard reciprocal-space methods.

In the case of insulators, where $f_i^{\vec{k}} = 1$ for all occupied bands,
the Wannier-like functions of (\ref{eqn:WannierLike}) are just the standard Wannier functions
$w_i^{\vec{R}}$ and thus are exponentially localized around the sites $\vec{R}$.\cite{WannierLocalization}
For the case of metals, where the occupations $f_i^{\vec{k}}$ are not constant, the $\bar{w}_i^{\vec{R}}$
are linear combinations of Wannier functions localized on different lattice sites
\begin{equation}
\bar{w}_i^{\vec{R}}(\vec{r}) = \sum_{\vec{R}'} w_i^{\vec{R}'}(\vec{r})
	\underbrace{
		\int \frac{\Omega d\vec{k}}{(2\pi)^3} e^{i\vec{k}\cdot(\vec{R}'-\vec{R})} \sqrt{f_i^{\vec{k}}}
	}_{F_i(\vec{R}'-\vec{R})},
\label{eqn:WannierLikeFunctions}
\end{equation}
with coefficients $F_i(\vec{R}'-\vec{R})$ given by a Fourier transform of the square-root
of the band occupation. For metals at zero temperature, the occupations are discontinuous
at a Fermi surface which leads to a polynomially decaying $F_i$.
In particular, for metals in three dimensions with a compact two-dimensional
Fermi surface, $F_i(\vec{R}) \sim R^{-2}$ for large $R$. At finite temperature $T$,
$F_i$ also decays exponentially with a length scale inversely proportional to $T$.
Consequently, the $\bar{w}_i^{\vec{R}}$ in metals are localized polynomially
at $T=0$ and exponentially with a temperature-dependent decay length at $T\ne 0$.
In all these cases, the localization of  $\bar{w}_i^{\vec{R}}$ closely mirrors
that of the one-particle density matrix.\cite{DensityMatrixLocalization}

Given the above properties of the Wannier-like functions $\bar{w}_i^{\vec{R}}$,
each pair density $\bar{\rho}^{\vec{R}}_{ij}(\vec{r})$ is localized
and diminishes with increasing $R$, since it is a product of
one function localized around $\vec{0}$ and another around $\vec{R}$.
The magnitude of the pair density decreases exponentially for insulators
and as $R^{-2}$ for metals at zero temperature, so that the corresponding
Coulomb self-energies decay exponentially and as $R^{-4}$ respectively.
The sum over unit cells in (\ref{eqn:ExchangeWannier}) converges in all these cases.

Next, to understand the convergence properties of actual calculations,
we repeat the above transformations with a finite $k$-point mesh with
$N_k$ samples instead of continuous integrals over the Brillouin zone.
A uniform $k$-point mesh centered on the $\Gamma$-point corresponds to
Kohn-Sham orbitals that are periodic on a supercell of volume $N_k\Omega$.
For uniform off-$\Gamma$ meshes, such as the Monkhorst-Pack grid,
the orbitals share a common Bloch phase on an $N_k\Omega$ supercell.
In all these cases, the pair densities $\bar{\rho}^{\vec{R}}_{ij}(\vec{r})$
are periodic on that $N_k\Omega$ supercell. Because of this periodicity, the expression
for the exchange energy (\ref{eqn:ExchangeWannier}) then remains unmodified except
that one of the integrals over space is restricted to a single $N_k\Omega$ supercell.
Converting the other integral over all of space to a sum over integrals restricted
to each $N_k\Omega$ supercell, and using the periodicity of the pair density,
\begin{equation}
E_X = \frac{-1}{2} \sum_{i,j,\vec{R}} \int_{N_k\Omega} d\vec{r} \int_{N_k\Omega} d\vec{r}'
	\bar{\rho}^{\vec{R}\ast}_{ij}(\vec{r}) K(\vec{r}-\vec{r}') \bar{\rho}^{\vec{R}}_{ij}(\vec{r}')
	,
\label{eqn:ExchangeWannierDiscrete}
\end{equation}
with
\begin{equation}
K(\vec{r}) = \sum_{\vec{S}}\frac{1}{|\vec{r}+\vec{S}|},
\label{eqn:Keff}
\end{equation}
where $\vec{S}$ are lattice vectors of the effective superlattice
of cell volume $N_k\Omega$ which arises from finite $k$-point sampling.
The summand in (\ref{eqn:Keff}) falls off only as $1/S$ causing $K(\vec{r})$
to diverge for all $\vec{r}$.  In fact, $K(\vec{r}) = \frac{1}{N_k\Omega}
\sum_{\vec{k},\vec{G}} e^{i(\vec{k}+\vec{G})\cdot\vec{r}} \frac{4\pi}{|\vec{k}+\vec{G}|^2}$,
so that it is again the $\vec{k}=\vec{G}=0$ component which needs special handling, as above.

Now consider a sufficiently dense $k$-point mesh such that the corresponding supercell
is much larger than the spatial extent of the localized $\bar{w}_i^{\vec{R}}$.
In that case, the periodic versions of $\bar{\rho}^{\vec{R}}_{ij}$ at finite
$k$-point sampling are identical (with exponentially small errors) in one,
appropriately centered, supercell to the original non-periodic localized ones.
Therefore, the contribution from the $\vec{S}=0$ term in $K(\vec{r})$ to
(\ref{eqn:ExchangeWannierDiscrete}), apart from errors decaying exponentially with the density
of the $k$-point mesh, is the true exchange energy of the infinite system (\ref{eqn:ExchangeWannier}).
The non-exponentially decaying errors in (\ref{eqn:ExchangeWannierDiscrete}) arise from the
contributions due to all the other super-cells $\vec{S}\neq 0$, which thus can be
eliminated completely by truncating the Coulomb potential so that $K(\vec{r}) = 1/r$!

Such truncation of the Coulomb potential on the Wigner-Seitz cell of the $k$-point sampled
superlattice with supercell volume $N_k\Omega$, in practice simply amounts to replacing
$\tilde{K}_{\vec{q}}$ in the standard reciprocal-space expression (\ref{eqn:ExchangePW})
by the Fourier transform of the truncated potential.
The minimum image convention (MIC) algorithm\cite{TruncationMIC} employs
\begin{flalign}
\tilde{K}\super{WS}_{\vec{q}} \approx
	\frac{4\pi}{q^2}\left(1-\exp\frac{-q^2}{4\alpha^2}\right)
	+ \frac{\Omega}{N_{\vec{r}}} \sum_{\vec{r}\in\textrm{WS}}
	e^{-i\vec{q}\cdot\vec{r}} \frac{\textrm{erf }\alpha r}{r},
\label{eqn:IsolatedKernelPreview}
\end{flalign}
and enables efficient construction of truncated kernels in the plane-wave basis.
The appendix derives this algorithm for arbitrary lattice systems as equation
(\ref{eqn:IsolatedKernel}); see appendix~\ref{sec:MIC} for a detailed explanation
of all the terms and approximations involved in (\ref{eqn:IsolatedKernelPreview}).
When such a truncated kernel replaces the periodic Coulomb kernel, the remaining error
in the exchange energy is due to the deviations of the periodic $\bar{\rho}^{\vec{R}}_{ij}$
from the infinite ones, within one supercell, as discussed above.
These deviations decay exponentially when the $\bar{w}_i^{\vec{R}}$ are exponentially localized,
leading to exponential convergence of the exchange energy with the number of $k$-points.

\begin{figure}
\includegraphics[width=\columnwidth]{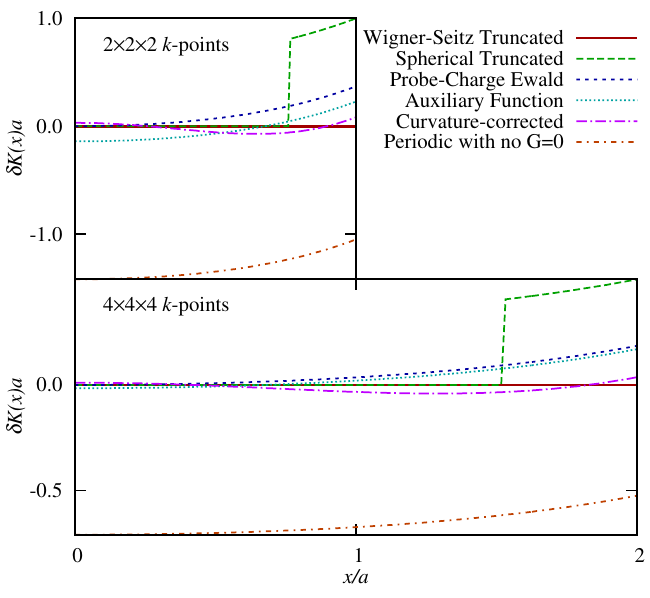}
\caption{Comparison of the discrepancy in the effective Coulomb Kernel
from $1/r$ of various methods for computing the Fock exchange energy.
This discrepancy is plotted along the $x$-direction from the origin to the boundary of
the cubic Wigner-Seitz cell of the $k$-point sampled super-lattice of a simple cubic lattice
of lattice constant $a$. Both axes have been scaled to be dimensionless.
The truncated Coulomb potentials have zero discrepancy for most or all of the supercell.
The auxiliary-function methods shift the periodic kernel to minimize the overall discrepancy,
while the probe-charge Ewald-sum method pins the discrepancy at the origin to 0.
\label{fig:DeltaK}}
\end{figure}

The general analysis we developed above not only allows us to establish the Wigner-Seitz super-cell
truncated potential as the natural method with ideal asymptotic convergence, but also provides
the framework for establishing analytically the convergence properties of other methods
\emph{a priori} by simply comparing the effective Coulomb kernels which they employ.
Ultimately, each reciprocal-space method prescribes a kernel $\tilde{K}(\vec{q})$
for use in (\ref{eqn:ExchangePW}), which translates to some $K(\vec{r})$
in the real space version (\ref{eqn:ExchangeWannierDiscrete}).
The error in any method compared to the ideal case is governed by the
discrepancy of $K(\vec{r})$ from $1/r$, given by
\begin{equation}
\delta K(\vec{r}) = \frac{1}{N_k\Omega} \sum_{\vec{k},\vec{G}}
	e^{i(\vec{k}+\vec{G})\cdot\vec{r}} \tilde{K}(\vec{k}+\vec{G}) - \frac{1}{r}
\end{equation}
for a particular $k$-point set.

FIG.~\ref{fig:DeltaK} compares the discrepancies in the effective Coulomb
kernels of various methods for a cubic lattice over a radial slice of
the effective supercell with different $k$-point meshes. $\delta K$ is exactly zero
within the domain of truncation for any truncated Coulomb potential, and hence
it is zero for the Wigner-Seitz truncated kernel in the entire supercell.
Truncation on a sphere of volume $N_k\Omega$, as proposed by Spencer and Alavi,\cite{SphericalTruncation}
achieves exponential convergence as well, but exhibits its asymptotic properties starting at larger $N_k$
since the sphere does not tile with the super-lattice and leads to some overlap between supercells.

The standard periodic kernel with $G=0$ projected out has a large discrepancy $\sim 1/L$,
where $L$ is the linear dimension of the supercell, and hence exhibits $\sim N_k^{-1/3}$
convergence. The auxiliary-function methods add a constant offset to the periodic kernel
by adjusting the $G=0$ component, which is chosen to minimize the discrepancy over the supercell
with some weight which depends on the choice of auxiliary function $f(\vec{q})$.
The curvature corrections proposed by Duchemin and Gygi\cite{AuxFunc-DucheminGygi}
adjust the constant offset in addition to the coefficients of $e^{i\delta\vec{k}\cdot\vec{r}}$
for the nearest-neighbors $\delta\vec{k}$ in the $k$-point mesh.
For the cubic supercell of length $L$ plotted in FIG.~\ref{fig:DeltaK},
this correction happens to be $(\cos(2\pi x/L)-1)/(\pi L)$ in the $x$ slice.
This additional freedom reduces the magnitude of the discrepancy more rapidly
with increasing $N_k$ than the case when only $G=0$ is adjusted.

An interesting alternative to the auxiliary-function method with similar accuracy
is the probe-charge Ewald-sum method.\cite{ProbeChargeEwald} Here, the $G=0$ component
of the Coulomb kernel is set to the potential at the origin from an array of unit negative
charges placed at all points of the $k$-point sampled super-lattice except the origin,
a Coulomb sum which can be computed readily using the Ewald method.\cite{Ewald3D} This amounts to adjusting
the discrepancy at the origin, $\delta K(0)$, to zero, as can be seen in FIG.~\ref{fig:DeltaK}.
Alternately, this method attempts to cancel the contributions from $\vec{S}\neq 0$ supercells
of (\ref{eqn:Keff}) in (\ref{eqn:ExchangeWannierDiscrete}) by neutralizing all those
supercells with a point charge. This would be exact if each $\bar{\rho}$ was spherically
symmetric, but in reality incurs an error asymptotically dominated by dipole-dipole
interactions $\sim L^{-3}$ and hence converges as $N_k^{-1}$ with $k$-point sampling.

The Wannier-function formalism presented here clearly establishes the
advantage of the truncated-potential methods and elucidates the asymptotic
convergence of the exchange energy computed using different methods.
Note that none of these methods require Wannier functions in practice;
each method prescribes a different replacement for the periodic plane-wave Coulomb kernel
$\tilde{K}_q = 4\pi/q^2$ in the standard reciprocal-space expression (\ref{eqn:ExchangePW}).
We compare the accuracy of these methods and demonstrate their analytically-predicted
asymptotic behavior for real materials in Section~\ref{sec:Results}.

\subsection{Extension to low-dimensional systems} \label{sec:PartiallyPeriodicExchange}

The preceding section shows how the integrable singularity in reciprocal-space calculations
of the exchange energy of systems with three-dimensional periodicity can be regularized
using auxiliary function methods or, ideally, by truncating the Coulomb potential to the
Wigner-Seitz cell simply by modifying the Fourier transform of the Coulomb kernel.
Reciprocal-space methods can also be applied to systems with lower-dimensional
periodicity by truncating the Coulomb potential along a subset of lattice directions.
The appendix details these types of truncations as well.
Specifically, the exchange energy in these geometries can be computed using
(\ref{eqn:ExchangePW}) by employing a partially truncated Coulomb potential
given by (\ref{eqn:SlabKernel}) or (\ref{eqn:WireKernel}) for $\tilde{K}_{\vec{q}}$,
and restricting the $k$-points sums to the two- or one-dimensional
Brillouin zone of the periodic directions alone.
The exchange energy still contains an integrable singularity, $q^{-1}$ for slabs and $\ln q$
for wires, that again needs to be addressed just as it did for bulk systems.
Before going on to present results in the next section, here, we briefly describe how we extend
each of the methods discussed above for bulk systems to the cases of these lower-dimensional geometries.

First, the auxiliary-function method for these geometries replaces the $G=0$ component
of the appropriate partially truncated Coulomb kernel to obtain
\begin{equation}
\tilde{K}_{\vec{q}}\super{aux} = 
\begin{cases}
	\tilde{K}_{\vec{q}}\super{slab/wire},& \vec{q}\neq 0 \\
	\Omega_\perp \left( 
		N_k \int_{\textrm{BZ}_\parallel} \frac{\Omega_\parallel d\vec{k}}{(2\pi)^d} f(\vec{k})
		- \sum_{\delta\vec{k}} f(\delta\vec{k})
	\right),& \vec{q}=0,
\end{cases}
\label{eqn:AuxFuncTruncated}
\end{equation}
where $\Omega_\parallel$ is the area/length of the two-/one-dimensional unit cell
in the periodic directions, BZ$_\parallel$ is the corresponding Brillouin zone and
$\Omega_\perp$ is the length/area of the artificial periodicity along the truncated directions.
The auxiliary function needs to match the singularity of the truncated Coulomb kernel,
and we adapt Carrier and coworkers' construction for arbitrary three-dimensional
lattices\cite{AuxFunc-Carrier} to lower dimensions. For a slab with lattice basis
vectors $\vec{a}_1$ and $\vec{a}_2$ in the periodic directions, and corresponding
reciprocal lattice vectors $\vec{b}_1$ and $\vec{b}_2$, the function
\begin{equation}
f(\vec{q}) = 2\pi^2 \left(\begin{array}{c}
	b_1^2 \sin^2\left(\vec{a}_1\cdot\frac{\vec{q}}{2}\right)
	+ b_2^2 \sin^2\left(\vec{a}_2\cdot\frac{\vec{q}}{2}\right) \\
	+ \frac{1}{2} \vec{b}_1\cdot\vec{b}_2 \sin(\vec{a}_1\cdot\vec{q})\sin(\vec{a}_2\cdot\vec{q})
\end{array}\right)^{-1/2}
\end{equation}
is periodic on the reciprocal lattice with a single singularity in the Brillouin zone
at $\vec{q}=0$ for arbitrary lattice vectors. Similarly,
\begin{equation}
f(\vec{q}) = -2\gamma + \ln\frac{a^2}{\sin^2(\vec{a}\cdot\vec{q}/2)}
\end{equation}
is a suitable auxiliary function for a wire with lattice vector $\vec{a}$
along the periodic direction, where $\gamma$ is the Euler-Mascheroni constant.
In this case, both the integral and the sum in (\ref{eqn:AuxFuncTruncated})
can be performed analytically to yield $\tilde{K}_0\super{aux}
= 2\Omega_\perp(\log2N_ka-\gamma)$.

Next, the probe-charge Ewald compensation method generalizes trivially to the
slab and wire geometries, and only requires the substitution of the usual
three-dimensional Ewald sum with the appropriate lower-dimensional analog.
(See Appendix~\ref{sec:Ewald} for details.) Interestingly, this method yields the same
$G=0$ component for the wire-geometry exchange kernel as the auxiliary function
method above, so that the two methods are identical for wires.

Finally, the Wigner-Seitz supercell truncated Coulomb potential requires no modification
for partially periodic systems. The truncation domain remains the Wigner-Seitz
cell of the $k$-point sampled super-lattice; one or two lattice directions
have only a single $k$-point and the boundaries of the supercell coincide
with the unit cell in those directions. These Wigner-Seitz cells become
increasingly anisotropic with increasing $N_k$ and spherical truncation
is no longer a viable option.

\section{Results} \label{sec:Results}

The analysis of Section~\ref{sec:ExchangeFormalism} identifies
Coulomb truncation with its asymptotic exponential convergence
as the natural choice for computing the exchange energy of periodic systems,
in comparison to auxiliary-function methods with polynomial convergence.
Here, we compare the accuracy of all these methods and demonstrate their
analytically-established asymptotic behavior for real materials
with a variety of electronic structures and dimensionalities.

Specifically, we consider four methods for computing the exact exchange energy,
the Wigner-Seitz truncated potential introduced here,
the spherical truncation of Spencer and Alavi,\cite{SphericalTruncation}
the probe-charge Ewald compensation method,\cite{ProbeChargeEwald}
and the auxiliary-function method with the general function applicable
to all lattice systems by Carrier and coworkers.\cite{AuxFunc-Carrier}
The first three of these methods trivially generalize to lower dimensions,
while the auxiliary-function method requires minor modifications
as detailed in Section~\ref{sec:PartiallyPeriodicExchange}.

We also compare the convergence of the exact exchange energy
using the above methods to that of the erf-screened exchange
employed in the range-separated HSE06 hybrid functional.\cite{HSE06}
In this functional, the Coulomb kernel in the non-local exchange energy
is replaced by the short-ranged $\textrm{erfc}(\omega r)/r$
with $\omega = 0.11 a_0^{-1}$, while the long-ranged part
is approximated using a semi-local functional.
The screened exchange avoids the $G=0$ singularity and the HSE06 functional
has so far achieved superior $k$-point convergence compared to regular
hybrid functionals with exact exchange.\cite{HSEconvergence}
Here, we demonstrate that employing Wigner-Seitz truncation
for the exact exchange energy puts the convergence of hybrid functionals
employing the exact non-local exchange energy (e.g. PBE0\cite{PBE0})
on par with that of the screened-exchange functionals (e.g. HSE06)
and even that of semi-local functionals employing no non-local
exchange whatsoever (e.g. PBE\cite{PBE}).

\subsection{Computational Details} \label{sec:CompDetails}

\begin{table}
\begin{center}\begin{tabular}{cc}
\hline\hline
Method & Compute Time [s] \\
\hline
Wigner-Seitz truncated & $555 \pm 11$ \\
Spherical truncated & $874 \pm 12$ \\
Auxiliary function & $552 \pm 11$ \\
Probe-charge Ewald & $543 \pm 10$ \\
\hline\hline
\end{tabular}\end{center}
\caption{Comparison of the average computation time for the exact exchange energy
using different regularization methods for hexagonal silicon carbide with
$8\times 8\times 8$ $k$-point sampling. The timing statistics are
from ten calculations for each method on identical 12-core Xeon compute nodes.
\label{tab:Timings}}
\end{table}

We have implemented all these methods in the open-source plane-wave density-functional
software JDFTx,\cite{JDFTx} where they are now publicly available. 
The specifics of these implementations are that the auxiliary-function and
probe-charge Ewald methods simply replace only the $\vec{G}=0$ value of the
$\vec{k}=\vec{k}'$ Coulomb kernel in (\ref{eqn:ExchangePW}) with a precomputed value.
The truncated potential methods, on the other hand, alter $\tilde{K}_{\vec{q}}$
in (\ref{eqn:ExchangePW}) for all $\vec{q}=\vec{G}+\vec{k}'-\vec{k}$.
Spherical truncation uses $\tilde{K}_{\vec{q}}$ defined analytically via (\ref{eqn:SphericalKernel}).
Wigner-Seitz truncation employs a precomputed kernel calculated by applying the MIC algorithm
(\ref{eqn:IsolatedKernel}) on the supercell, as detailed in the appendix,
and then redistributing the resulting supercell kernel to unit cell kernels for each $\vec{k}'-\vec{k}$.

TABLE~\ref{tab:Timings} shows that the computational overhead for looking up the precomputed kernel
in the Wigner-Seitz truncated method is negligible, and results in compute times equal to
the auxiliary function and probe-charge Ewald methods, within run-to-run variations.
In fact, this overhead is negligible compared to that of the extra transcendental
(cosine) evaluations in spherical truncation; precomputing the kernel also optimizes
spherical truncation and we report the analytical evaluation time here only to illustrate
the negligible lookup overhead. Next, the computational effort to calculate and predistribute
the kernel is negligible compared to a single evaluation of the exchange energy:
a mere 1.4 s for the example of TABLE~\ref{tab:Timings}.
Finally, the memory overhead of the precomputed kernel is comparable to
four Kohn-Sham bands, and is therefore negligible for most systems.

In order to study a large number of materials and $k$-point configurations
within the available computational resources, rather than performing fully
self-consistent calculations, we first determine converged Kohn-Sham orbitals of a
density-functional calculation using the semi-local PBE exchange and correlation functional,\cite{PBE}
and then compute the exchange energies from these orbitals according to the above methods.
The calculations employ norm-conserving pseudopotentials at a kinetic energy
cutoff of 30~$E_h$. TABLE~\ref{tab:Parameters} summarizes the
unit-cell parameters for the systems studied below.

\begin{table}
\begin{center}\begin{tabular}{ccccccc}
\hline\hline
System          & Unit cell & $N_a$ & $a$ [\AA] & $c$ [\AA] &       Ref.              & FIG. \\
\hline
2H-SiC          & Hexagonal  & 4  & 3.076  & 5.048  & \citenum{Lattice-AlphaSiC} & \ref{fig:AlphaSiC},\ref{fig:TotalE}(b) \\
3C-SiC          & FCC        & 2  & 4.3596 &   -    & \citenum{Lattice-BetaSiC}  & \ref{fig:DiamondStructure}(b) \\
Ice XIc         & BCT$^a$    & 6  & 4.385  & 6.219  & \citenum{IceGroundState}   & \ref{fig:Ice} \\
Si              & FCC        & 2  & 5.431  &   -    & \citenum{CRC-Handbook}     & \ref{fig:DiamondStructure}(a),\ref{fig:TotalE}(a) \\
Platinum        & FCC        & 1  & 3.924  &   -    & \citenum{CRC-Handbook}     & \ref{fig:Pt},\ref{fig:TotalE}(c) \\
Diamond         & FCC        & 2  & 3.567  &   -    & \citenum{CRC-Handbook}     & \ref{fig:DiamondStructure}(c) \\
Graphite        & Hexagonal  & 4  & 2.461  & 6.709  & \citenum{CRC-Handbook}     & \ref{fig:Graphite} \\
Graphene        & Hexagonal  & 2  & 2.46   & 10$^b$ & PBE$^c$                    & \ref{fig:Graphene} \\
(8,0) SWCNT$^d$ & Tetragonal & 32 & 25$^b$ & 4.32   & PBE$^c$                    & \ref{fig:Nanotube} \\
\hline\hline
\end{tabular}\end{center}
\begin{itemize}
\item[$^a$] {\small Body-centered tetragonal.}
\item[$^b$] {\small Coulomb potential is truncated along these directions.}
\item[$^c$] {\small DFT lattice constants using the PBE functional.\cite{PBE}}
\item[$^d$] {\small Single-walled carbon nanotube.}
\end{itemize}
\caption{Unit-cell parameters for the systems studied here, including citations for
experimental lattice constants and references to figures with corresponding results.
$N_a$ is the number of atoms in the primitive unit cell of each calculation.
\label{tab:Parameters}}
\end{table}

Figures~\ref{fig:DiamondStructure}-\ref{fig:Nanotube} show the deviation of
the exact and screened exchange energies at finite $k$-point configurations
from their $k$-point-converged values for a variety of systems.
The left-hand panels show this deviation for coarse $k$-point meshes on a linear scale,
while the right-hand panels illustrate the asymptotic convergence on a logarithmic energy scale.
The base line of $k$-points is insufficient to reliably fit power laws and the dotted lines
are only a visual guide with the expected exponent for polynomial convergence.
For the three-dimensional systems, we study both isotropic and anisotropic $k$-point meshes.
The plots explicitly label anisotropic $k$-point configurations, whereas the unlabeled points
at integer values of $X \equiv N_k^{1/3}$ correspond to $X{\times}X{\times}X$ $k$-point meshes.

\subsection{Insulators}

\begin{figure}
\includegraphics[width=\columnwidth]{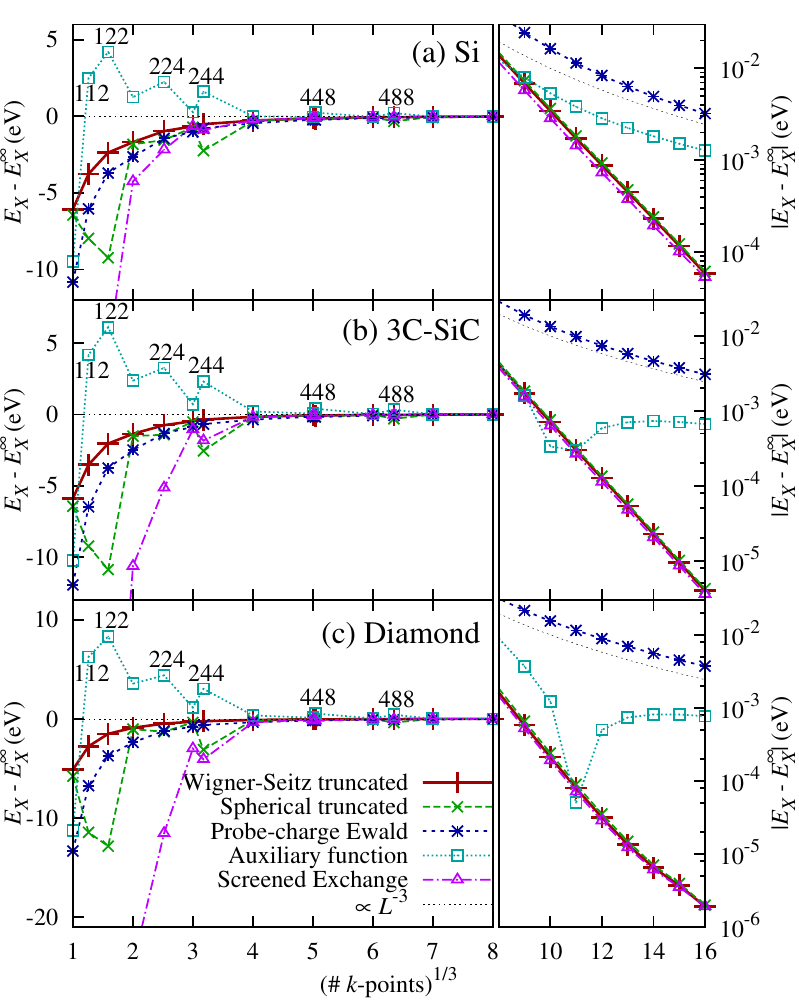}
\caption{Convergence of exact and screened exchange energies for three
semiconductors and insulators in the diamond (zinc-blende) structure:
(a) silicon, (b) cubic silicon carbide (phase 3C), and (c) diamond.
See last paragraph of Section~\ref{sec:CompDetails} for details.
The non-monotonicity in the absolute asymptotic error of the
auxiliary-function results is due to a change in sign of that error.
\label{fig:DiamondStructure}}
\end{figure}

We begin our computational study with a sequence of semiconductors and insulators
in the high-symmetry diamond structure. FIG.~\ref{fig:DiamondStructure}
compares the deviation of the exchange energy of silicon, cubic silicon carbide and diamond
at various finite $k$-point configurations from the infinite limit
for different singularity regularization methods. 

While the order of magnitude of error in the exact exchange energy with coarse
$k$-point meshes is comparable for all methods, Wigner-Seitz truncation
typically yields significantly lower errors than do the other methods.
The Wigner-Seitz truncated and the probe-charge Ewald methods
(red $+$'s and blue $\plustimes$'s respectively in FIG.~\ref{fig:DiamondStructure})
exhibit smooth convergence for all $k$-point meshes including anisotropic ones,
whereas the remaining methods incur higher errors for anisotropic $k$-point meshes.
The pattern of errors with $k$-points for each method is similar for the three
materials with the same underlying Bravais lattice and point-group symmetries.

In contrast, the asymptotic exponential convergence of the truncated methods leads to
orders of magnitude reduction in error for fine $k$-point meshes, compared to
the probe-charge Ewald and auxiliary-function methods, which exhibit $L^{-3}$ convergence.
The exponential decay length of the error in the exchange energy with respect to $L$,
taken here to be the nearest neighbor distance in the effective $k$-point sampled super-lattice,
decreases from $5.5$~\AA~in silicon through $3.5$~\AA~in cubic silicon carbide to $2.5$~\AA~in diamond.
The corresponding band gaps $\Delta$ are $1.1$~eV, $2.3$~eV and $5.5$~eV respectively.
The decay length varies roughly as $\Delta^{-1/2}$, similar to the density-matrix
localization length scale of tight-binding insulators.\cite{DensityMatrixLocalization}
Consequently, the relative accuracy of the truncated methods for insulators
increases dramatically with increasing band gap as seen in FIG.~\ref{fig:DiamondStructure}.
Note that the accuracy of the truncated potential methods for the exact exchange energy
matches that of the screened exchange energy (red $+$'s and green $\times$'s
versus pink $\Delta$'s), indicating that these methods are truly limited only
by Brillouin-zone discretization errors in the underlying Kohn-Sham orbitals.

\begin{figure}
\includegraphics[width=\columnwidth]{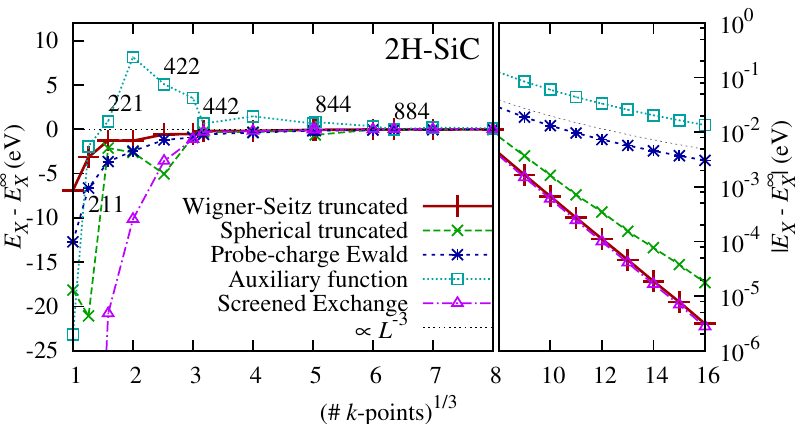}
\caption{Convergence of exact and screened exchange energies for hexagonal silicon carbide (phase 2H).
See last paragraph of Section~\ref{sec:CompDetails} for details.
\label{fig:AlphaSiC}}
\end{figure}

In high-symmetry materials, the accuracy of spherical truncation is similar
to Wigner-Seitz truncation for isotropic $k$-point meshes since the
Wigner-Seitz cell of the effective super-lattice is approximately spherical
(rhombic dodecahedron for the FCC unit cell in the zinc-blende structure). 
This is no longer the case for lower symmetry crystals such as hexagonal silicon
carbide (phase 2H with the wurtzite structure) shown in FIG.~\ref{fig:AlphaSiC}.
In this anisotropic case, the accuracy of the Wigner-Seitz truncated method (red $+$'s)
continues to match that of screened exchange (pink $\Delta$'s).
On the other hand, spherical truncation (green $\times$'s), although still
exponentially convergent, is an order of magnitude less accurate.
Similarly, amongst the asymptotically $L^{-3}$ convergent methods,
the superior accuracy of the probe-charge Ewald method for anisotropic $k$-point
meshes in the high-symmetry crystal carries forward to superior accuracy
overall for lower-symmetry crystals, in comparison to the auxiliary-function method.

\begin{figure}
\includegraphics[width=\columnwidth]{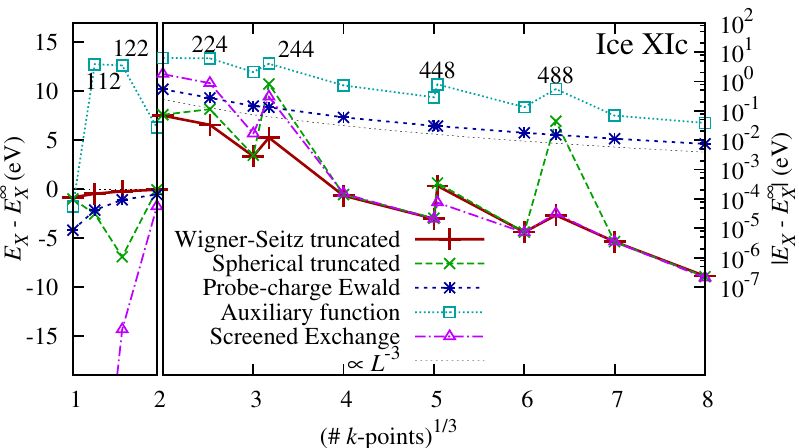}
\caption{Convergence of exact and screened exchange energies for proton-ordered cubic ice.
See last paragraph of Section~\ref{sec:CompDetails} for details.
\label{fig:Ice}}
\end{figure}

The differences between the methods are most dramatic for proton-ordered cubic ice XIc
(the proposed ground state structure\cite{IceGroundState}) shown in FIG.~\ref{fig:Ice}.
The highly-localized states in this material cause dramatic improvements in
accuracy for the truncated methods even for coarse $k$-point meshes.
Once again, lowered symmetry significantly favors the probe-charge
Ewald method in comparison to the auxiliary-function method (blue $\plustimes$'s versus cyan $\square$'s).

\subsection{Metals}

\begin{figure}
\includegraphics[width=\columnwidth]{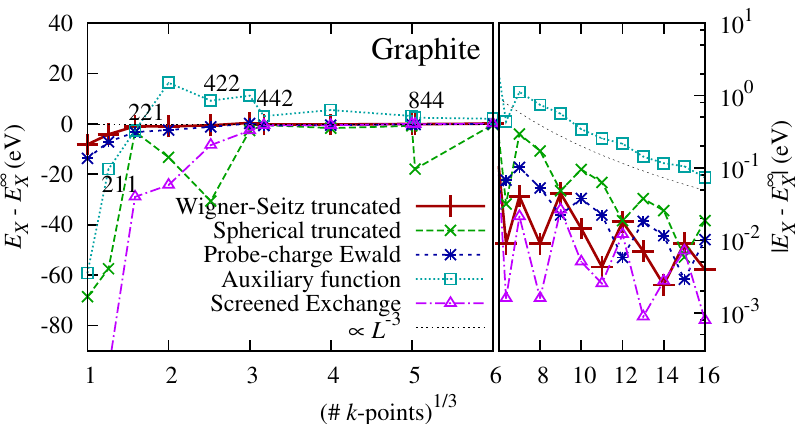}
\caption{Convergence of exact and screened exchange energies for graphite.
See last paragraph of Section~\ref{sec:CompDetails} for details.
\label{fig:Graphite}}
\end{figure}

As demonstrated above, the exponential localization of Wannier functions leads
directly to exponential convergence in the case of truncated Coulomb interaction methods.
In contrast, we expect the discontinuity at the Fermi surface at zero temperature
to lead to algebraic convergence in metallic systems, which we explore now.

FIG. \ref{fig:Graphite} shows the convergence behavior of the various
methods for the case of graphite, which is semi-metallic. Once again,
the probe-charge Ewald and auxiliary-function methods exhibit $L^{-3}$ convergence.
Wigner-Seitz truncation no longer exhibits exponential convergence,
but remains the most accurate method for computing the exchange energy,
with accuracy comparable to that of screened exchange as before.
Spherical truncation and the auxiliary-function method are less
accurate due to the lower symmetry of the crystal structure in this case.

\begin{figure}
\includegraphics[width=\columnwidth]{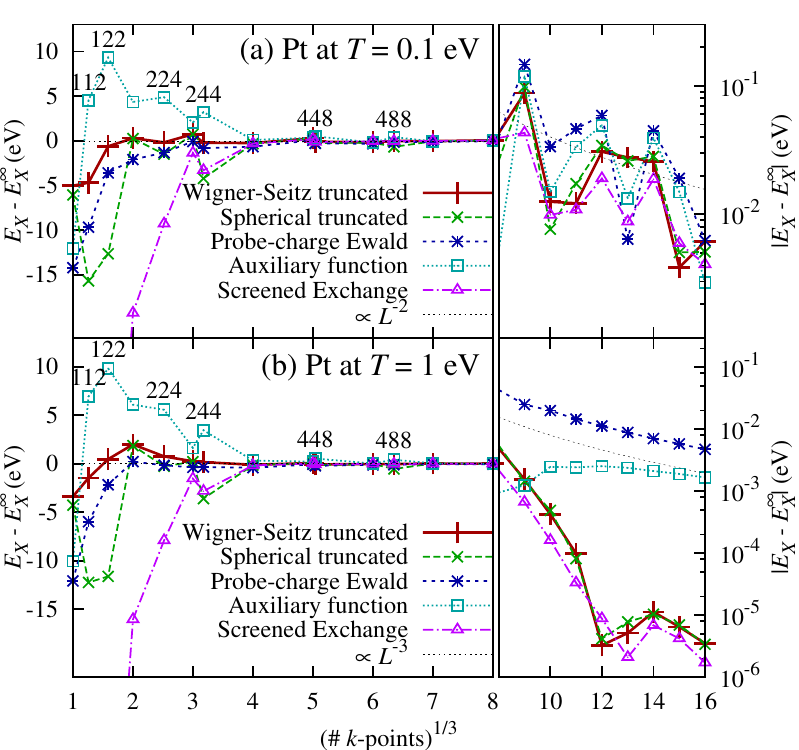}
\caption{Convergence of exact and screened exchange energies for face-centered cubic
metallic platinum at electron temperatures, (a) $T=0.1$~eV and (b) $T=1$~eV.
See last paragraph of Section~\ref{sec:CompDetails} for details.
The non-monotonicity in the exponentially-convergent results in (b)
for exact exchange with truncated-potentials and screened exchange
is due to a change in sign of the error near $N_k^{1/3} \sim 12$.
\label{fig:Pt}}
\end{figure}

The exponents governing the localization in graphite are complicated by the
layered quasi-two-dimensional structure with weak inter-planar coupling.
We analyze those details in the related two-dimensional
material graphene in Section~\ref{sec:ResultsLowerDimensions},
and here now focus on a simpler, three-dimensional metal, platinum.

In simple metals, the Wannier-like functions $\bar{w}_i^{\vec{R}}(\vec{r})$
given by (\ref{eqn:WannierLikeFunctions}) decay $\sim r^{-2}$,
as discussed in Section~\ref{sec:RealSpaceAnalysis}.
This leads to $r^{-4}$ decay of the pair densities and consequently
$\sim \int_L^\infty 4\pi r^2 dr (1/r)r^{-4} \sim L^{-2}$ errors due to
truncation in the Coulomb self energies in (\ref{eqn:ExchangeWannierDiscrete}).
The $L^{-2}$ errors dominate the asymptotic convergence of all the methods
for metals at low temperatures, as shown in FIG.~\ref{fig:Pt}(a) for platinum.
However, Wigner-Seitz truncation (red $+$'s) continues to yield the highest accuracy
for exact exchange in practice, particularly for coarse $k$-point meshes.

At finite Fermi temperature $T$ for the electrons, the exponential decay length
scales as $at/T$, where $a$ is the lattice constant and $t$ is the typical band width.
When the number of $k$-points along each dimension exceeds approximately $t/T$,
this decay length plays an analogous role to the Wannier-function length scale of insulators.
FIG.~\ref{fig:Pt}(b) shows the restored exponential convergence of the truncated methods
and screened exchange, and the $L^{-3}$ convergence of the other methods, in this regime.

\subsection{Lower dimensional materials} \label{sec:ResultsLowerDimensions}

\begin{figure}
\includegraphics[width=\columnwidth]{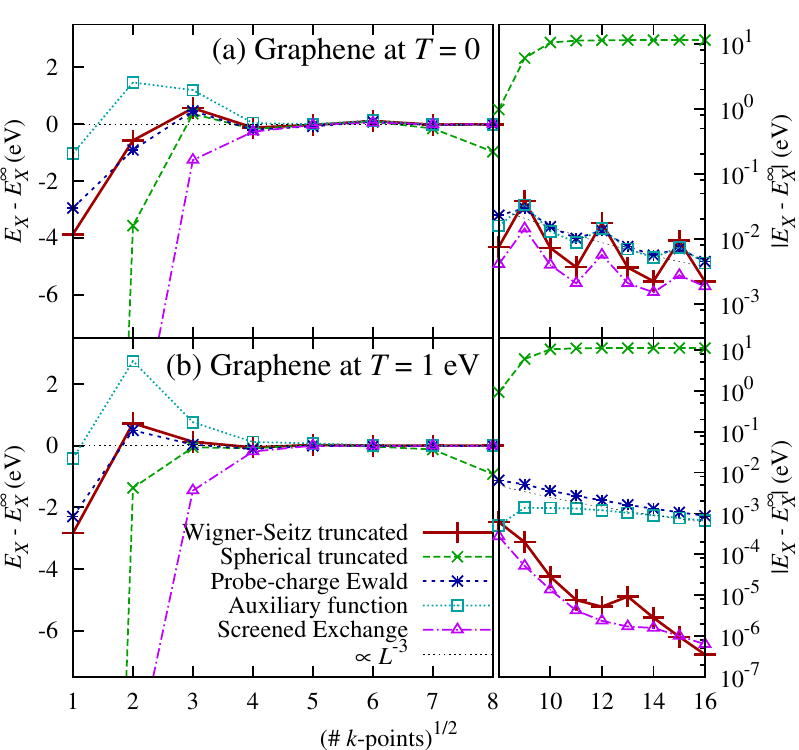}
\caption{Convergence of exact and screened exchange energies for graphene at
electron temperatures, (a) $T=0$ and (b) $T=1$~eV.
See last paragraph of Section~\ref{sec:CompDetails} for details.
The non-monotonicity in the exponentially-convergent results in (b) for Wigner-Seitz
truncated exact exchange is due to a change in sign of the error near $N_k^{1/2} \sim 11$.
\label{fig:Graphene}}
\end{figure}

In lower dimensional semiconducting or insulating systems, we should expect the
localization of the underlying Wannier functions to allow for exponential $k$-point
convergence with an appropriately chosen method from Section~\ref{sec:PartiallyPeriodicExchange}.
For the metallic cases, the reduced dimensionality can lead to different
exponents for the polynomial convergence, which we also explore here.

The semi-metallic behavior of the two-dimensional material graphene is particularly interesting.
In this case, the localization properties are determined by the phase twist of Bloch
functions in $k$-space about the Dirac point. The two-dimensional Fourier transform
of that phase twist yields an $r^{-2}$ decay of the Wannier-like functions.
The pair densities fall off as $r^{-4}$ leading to truncation errors
in the Coulomb self energies in (\ref{eqn:ExchangeWannierDiscrete})
that scale as $\int_L^\infty 2\pi r dr (1/r) r^{-4} \sim L^{-3}$.
FIG.~\ref{fig:Graphene}(a) shows that all methods, therefore, exhibit
$L^{-3}$ asymptotic convergence for graphene at zero temperature.
The errors oscillate with a period of 3 $k$-points per dimension because the discrete
$k$-point mesh includes the special Dirac point when the sampling is a multiple of 3.

At sufficiently high temperatures, the exponential length scale $at/T$ of the Wannier-like
functions becomes relevant at practical $k$-point meshes, just as in three-dimensional metals.
FIG.~\ref{fig:Graphene}(b) shows that this length scale restores the exponential convergence
of the Wigner-Seitz truncated method and screened exchange (red $+$'s and pink $\Delta$'s respectively).

Truncation on a three-dimensional sphere is no longer meaningful in these lower
dimensional materials, and it gives reasonable results only for intermediate
$k$-point meshes which minimize the aspect ratio of the supercell.
Analytically Fourier transforming the Coulomb potential truncated on the appropriate
`lower-dimensional spheres', finite cylinders in two-dimensional materials and
finite right prisms in one-dimensional materials, is no longer possible.
Wigner-Seitz truncation using the MIC algorithm (\ref{eqn:IsolatedKernel})
is clearly the method of choice in these geometries.

\begin{figure}
\includegraphics[width=\columnwidth]{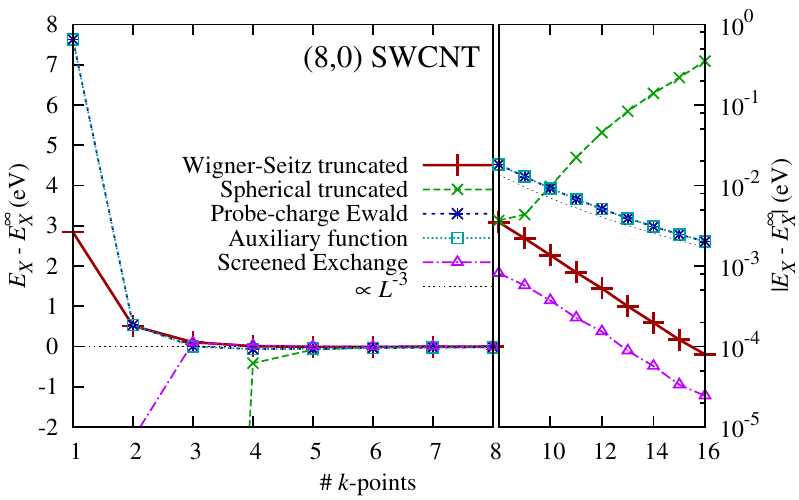}
\caption{Convergence of exact and screened exchange energies for the
semiconducting (8,0) single-walled carbon nanotube.
See last paragraph of Section~\ref{sec:CompDetails} for details.
\label{fig:Nanotube}}
\end{figure}

Finally, in one dimensional systems, the probe-charge Ewald and auxiliary-function
methods are identical as proved in Section~\ref{sec:PartiallyPeriodicExchange}.
As FIG.~\ref{fig:Nanotube} shows, for a semiconducting (8,0) single-walled
carbon nanotube (SWCNT), both of these methods yield rather poor $L^{-3}$ convergence,
in contrast to the exponential convergence of the Wigner-Seitz truncation
and the screened exchange interaction for this system.

\subsection{Total Energy Convergence}

\begin{figure}
\includegraphics[width=\columnwidth]{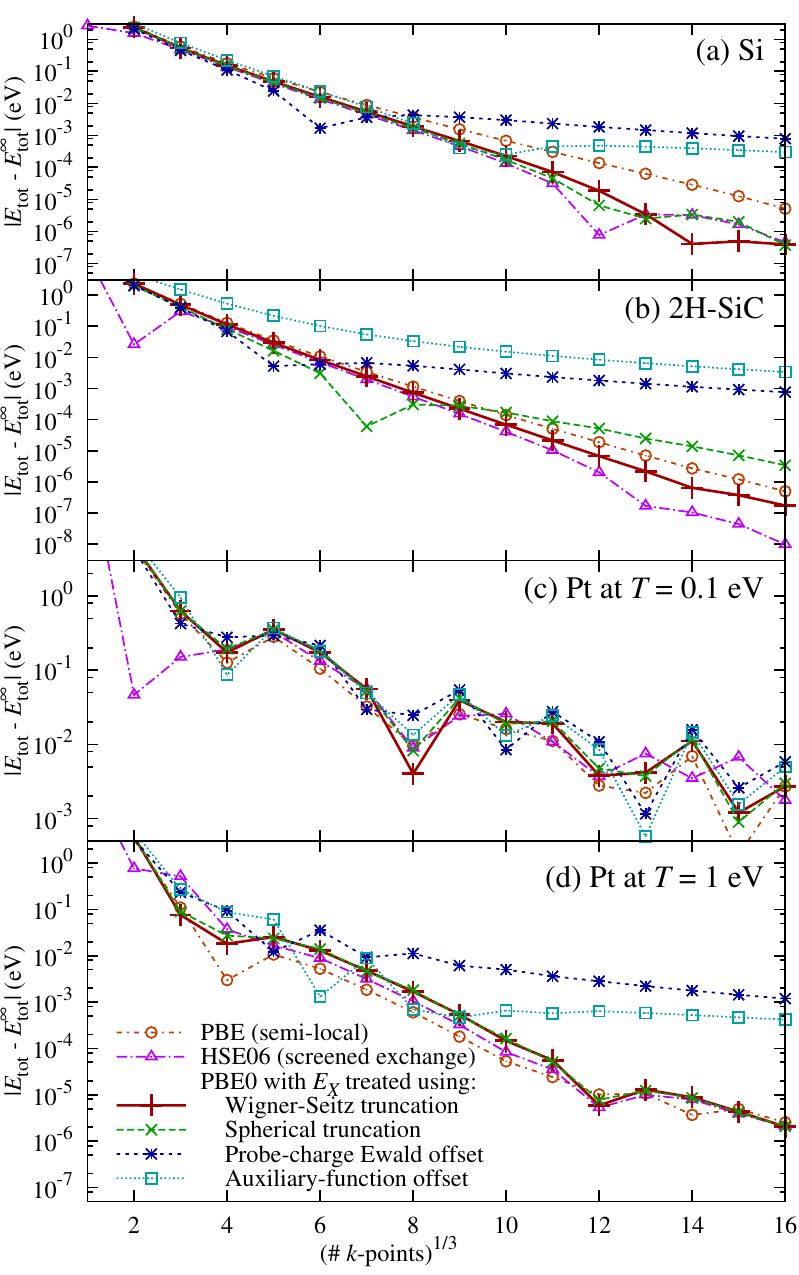}
\caption{Total energy convergence of the semi-local functional PBE and the screened-exchange
hybrid functional HSE06 compared to the hybrid functional PBE0 with exact exchange computed
using various methods for (a) silicon (b) hexagonal silicon carbide (phase 2H)
and metallic platinum at (c) $T=0.1$~eV and (d) $T=1$~eV.
The logarithmic energy scale shows the deviation of the energy for each functional
at finite $k$-point meshes from the converged energy for the same functional.
The non-monotonicity in the exponentially-convergent results in (d)
for hybrid functionals with truncated-potential exact exchange and screened exchange
is due to a change in sign of the error near $N_k^{1/3} \sim 12$.
\label{fig:TotalE}}
\end{figure}

The results in the preceding sections demonstrate that calculation of exact exchange
with Wigner-Seitz truncation, with relatively few exceptions, generally requires
fewer $k$-points to reach a given level of convergence than all other methods.
Moreover, we have seen that calculation of the long-ranged exact exchange, when performed with the
Wigner-Seitz truncated method, competes with the short-ranged screened exchange of the HSE06 functional,
which models the long-ranged components of the exchange energy within a semi-local approximation.
We now ask whether the Wigner-Seitz method makes it possible to evaluate exact-exchange functionals
on the same, relatively modest $k$-point meshes needed for simple density-functional theory calculations.

To address the above issues, FIG.~\ref{fig:TotalE} compares the total energy convergence of a purely
semi-local density functional (PBE\cite{PBE}), a standard hybrid functional (PBE0\cite{PBE0})
employing exact exchange computed using various standard approaches as well as our Wigner-Seitz approach,
and a hybrid functional (HSE06\cite{HSE06}) employing short-ranged screened exchange.
The results in FIG.~\ref{fig:TotalE} show that the total energy convergence of
the exact-exchange hybrid functional PBE0, when computed using truncated potentials,
is indeed comparable to that of the screened-exchange hybrid functional HSE06.
Moreover, computing exact exchange with truncated methods not only matches
the convergence of traditional semi-local density functionals, but sometimes even
outperforms their convergence for insulators (as in FIG.~\ref{fig:TotalE}a,b).

Spherical truncation yields similar convergence to Wigner-Seitz truncation for
high-symmetry insulators (green $\times$'s versus red $+$'s in FIG.\ref{fig:TotalE}a),
and is less accurate for lower symmetry ones (FIG.~\ref{fig:TotalE}b), as expected.
The auxiliary function and probe-charge Ewald methods limit the total energy convergence
of PBE0 to $L^{-3}$ (cyan $\square$'s and blue $\plustimes$'s in FIG.~\ref{fig:TotalE}(a,b,d)),
in contrast to the exponential convergence of PBE and HSE06.

The $r^{-2}$ localization of the Wannier-like functions in metals at low temperature limit
the convergence of all methods, including the semi-local functionals (FIG.~\ref{fig:TotalE}c).
The $L^{-2}$ errors in the tails of the Kohn-Sham orbitals lead to $L^{-2}$
errors in the Kohn-Sham kinetic energy, which dominate in this situation.
Consequently, all methods for treating exchange yield total-energy
convergence of PBE0 similar to that of PBE and HSE06 for low-temperature metals,
with Wigner-Seitz truncation only marginally better than the others.

At sufficiently high temperatures (FIG.~\ref{fig:TotalE}d), the exponential
length scale $at/T$ becomes relevant at practical $k$-point configurations as before,
leading to exponential total-energy convergence of the semi-local functionals.
As in the case of insulators, the auxiliary-function methods limit the accuracy
of the hybrid functional while the truncated-potential methods yield total-energy
convergence for the exact-exchange hybrid functional on par with the
screened-exchange and traditional semi-local functionals.

\section{Conclusions}

Hybrid density-functionals enable high-accuracy predictions within Kohn-Sham theory,
but state-of-the-art methods for computing such functionals necessitate dense
$k$-point meshes in calculations of periodic systems due to the regularization methods
commonly employed for the singular Coulomb integrals in the exact-exchange energy. 
In this work, we analyze the exchange energy in a real-space formalism based on
Wannier functions and show that the dominant errors at finite $k$-point meshes
arise from the resultant artificial periodicity of the Wannier functions.
Truncating the Coulomb potential on the Wigner-Seitz cell of the effectively-sampled
super-lattice is therefore the ideal method to minimize these errors.

We then prove analytically that the exchange energy computed using truncated potentials
converges exponentially with the number of $k$-points, $N_k$, whenever
the one-particle density matrix is exponentially localized.
This is the case for all but metallic systems at $T=0$.
In contrast, the frequently-employed auxiliary-function regularization methods
for the exchange kernel suffer from a dipole-dipole interaction with
the artificial images that limits the asymptotic convergence to $N_k^{-1}$.

To deliver this exponential asymptotic convergence to practical calculations,
we then develop in the appendix, general and efficient constructions for the
Wigner-Seitz truncated Coulomb potentials in the plane-wave basis.
Additionally, we generalize the truncated-potential and auxiliary-function
methods to slab-like or wire-like systems with lower-dimensional periodicity by
using Coulomb potentials truncated on a subset of lattice directions.

Finally, we explore the accuracy of several methods for computing exchange energies
applied to materials with varying electronic structure and dimensionality.
The Wigner-Seitz truncation motivated in this work systematically yields the
most accurate results for all of these systems, demonstrates exponential
convergence with Brillouin-zone sampling for all but metals at zero temperature,
and, most importantly, delivers $k$-point convergence for hybrid functionals
on par with that of screened-exchange and even traditional semi-local functionals.
The findings of this work will enable accurate calculations of periodic systems employing
exact-exchange hybrid functionals at the same computational effort as those employing
screened-exchange functionals, and bring us one step closer to widespread \emph{ab initio}
studies of processes such as catalysis at solid surfaces that require chemical accuracy.

We thank K. A. Schwarz and D. Gunceler for their helpful suggestions in improving this manuscript.
This work was supported as a part of the Energy Materials Center at Cornell (EMC$^2$),
an Energy Frontier Research Center funded by the U.S. Department of Energy,
Office of Science, Office of Basic Energy Sciences under Award Number DE-SC0001086.

\appendix
\section{Truncated Coulomb potentials}

Computations for periodic systems frequently employ the plane-wave basis,\cite{DFT-review}
which presents the advantage of systematic and exponential basis-set convergence controlled
by a single parameter, namely the kinetic energy cutoff, or equivalently, the Nyquist frequency.
This advantage can be extended to non-periodic systems and systems with lower-dimensional periodicity
such as slabs and wires by using truncated Coulomb potentials.\cite{TruncationAnalytic,TruncationWS}
Additionally, above, truncated Coulomb potentials proved particularly useful in
the computation of the exchange energy even for periodic systems.

As shown above, truncation of the potential on the Wigner-Seitz cell leads to
the most accurate method for the exchange energy, and it is also the most efficient method
for lower-dimensional geometries since, as discussed by Ismail-Beigi,\cite{TruncationWS}
it localizes the $G=0$ singularity of the Coulomb kernel to a single point.
However, the singular Fourier integrals required to construct plane-wave
Wigner-Seitz truncated Coulomb kernels cannot be solved analytically in general,
and are prohibitively expensive to compute numerically.
Here, we develop a general and efficient $\mathcal{O}(N\ln N)$ construction
for these kernels based on the minimum-image convention (MIC) method.\cite{TruncationMIC}
Finally, as a practical matter, we note that determining the Coulomb kernel is a
one-time computation because the corresponding storage requirements are modest.

To establish our normalization conventions for the plane-wave basis,
we expand periodic charge densities as $\rho(\vec{r}) =
\sum_{\vec{G}} e^{i\vec{G}\cdot\vec{r}} \tilde{\rho}_{\vec{G}}$
where $\vec{G}$ are reciprocal lattice vectors for the unit cell of volume $\Omega$.
The interaction energy under a translationally invariant potential
\begin{flalign}
U_{12} &= \int_\Omega d\vec{r}_1 \int d\vec{r}_2 \rho_1^\ast(\vec{r}_1)
	K(\vec{r}_1-\vec{r}_2)\rho_2(\vec{r}_2) \nonumber\\
&= \Omega\sum_{\vec{G}}\tilde{\rho}^\ast_{1\vec{G}}
	\tilde{\rho}_{2\vec{G}} \tilde{K}_{\vec{G}}
\label{eqn:CoulombEnergyPW}
\end{flalign}
is then diagonal in reciprocal space, with
$\tilde{K}_{\vec{G}} = \int d\vec{r} e^{-i\vec{G}\cdot\vec{r}} K(r)$.
Here, we denote integrals over unit cells by $\int_\Omega d\vec{r}$
and integrals over all space by $\int d\vec{r}$.
For the long-ranged Coulomb interaction $K(r) = 1/r$, the plane-wave kernel
$\tilde{K}_{\vec{G}} = 4\pi/G^2$ is singular at $\vec{G}=0$, and the
interaction energy is finite only for neutral unit cells.
In practice, the interaction energies of charged subsystems are
computed by excluding the $\vec{G}=0$ term, which amounts to
adding a uniform neutralizing background charge to each subsystem.

Treating non-periodic systems in the plane-wave basis requires
the elimination of interactions between different unit cells
in a translationally-invariant manner in order to preserve
the efficiency and accuracy of the Fourier spectral method.
If the Coulomb potential is truncated so that it is zero outside the
first Wigner-Seitz cell and the charge densities are confined to half
the domain of truncation, then the interaction within each unit cell
remains unmodified and the interaction between unit cells is exactly zero.
The simplest choice for accomplishing this truncates the Coulomb potential
outside a sphere of radius $R$ smaller than the in-radius of the
Wigner-Seitz cell, leading to the analytical plane-wave kernel,
\begin{equation}
\tilde{K}\super{sph}_{\vec{G}} = \int_{|\vec{r}|<R} d\vec{r} e^{-i\vec{G}\cdot\vec{r}} \frac{1}{r}
 =	\begin{cases}
		2\pi R^2, & \vec{G}=0 \\
		\frac{4\pi}{G^2}(1-\cos GR), & \vec{G}\neq 0.
	\end{cases}
\label{eqn:SphericalKernel}
\end{equation}

\subsection{Minimum-image convention (MIC) method} \label{sec:MIC}

From the above arguments, the natural choice for the truncated potential is clearly
\begin{equation}
\tilde{K}\super{WS}_{\vec{G}} = \int\sub{WS} d\vec{r} e^{-i\vec{G}\cdot\vec{r}} \frac{1}{r},
\end{equation}
where $\int\sub{WS}$ represents integration over the first Wigner-Seitz cell.
However, the singularity at the origin and the polyhedral domain of integration
preclude general analytical solutions and straightforward numerical quadratures.
Martyna and Tuckerman introduced an approximate construction for Coulomb potentials
truncated on parallelepiped domains based on range-separation techniques.
Here, we generalize this so-called minimum-image convention (MIC)
method\cite{TruncationMIC} to Wigner-Seitz cells of arbitrary lattice systems.

Employing a range-separation parameter $\alpha$, we approximate this kernel by
\begin{flalign}
\tilde{K}\super{WS}_{\vec{G}} &= \int\sub{WS} d\vec{r} e^{-i\vec{G}\cdot\vec{r}}
	\left( \frac{\textrm{erfc }\alpha r}{r} + \frac{\textrm{erf }\alpha r}{r} \right) \nonumber\\
&\approx \frac{4\pi}{G^2}\left(1-\exp\frac{-G^2}{4\alpha^2}\right)
	+ \frac{\Omega}{N_{\vec{r}}} \sum_{\vec{r}\in\textrm{WS}}
	e^{-i\vec{G}\cdot\vec{r}} \frac{\textrm{erf}\alpha r}{r},
\label{eqn:IsolatedKernel}
\end{flalign}
where the discrete sum over $\vec{r}$ is a quadrature on the
Wigner-Seitz cell with $N_{\vec{r}}$ nodes as described below.
The short-ranged first term is localized to the Wigner-Seitz cell
by choice of $\alpha$, so that it is unaffected by the truncation
and can be evaluated analytically.
The error in this term can be reduced to machine precision $\epsilon$
by choosing $\alpha = \sqrt{-\ln\epsilon}/R\sub{in}$,
where $R\sub{in}$ is the in-radius of the Wigner-Seitz cell.
The $\vec{G}=0$ component is well-defined due to the finite real-space range,
and is understood to be equal to its $G\to0$ limit given by $\pi/\alpha^2$.

The second term of (\ref{eqn:IsolatedKernel}) has a long-ranged
smooth integrand and is approximated by a Gauss-Fourier quadrature,
evaluated as a fast Fourier transform (FFT) in practice.
This quadrature consists of nodes on a uniform parallelepiped mesh
with uniform weights, which we remap using the periodicity of the
lattice to the first Wigner-Seitz cell.
In the interior of the integration domain, the integrand
is bandwidth-limited as $\exp(-G^2/4\alpha^2)$,
and the error in the Fourier quadrature can be reduced to $\epsilon$
by choosing an FFT resolution such that the Nyquist frequency exceeds
$2\alpha\sqrt{-\ln\epsilon} = -2\ln\epsilon/R\sub{in}$.
The cusps in the periodic repetition of the integrand at the boundaries
of the Wigner-Seitz cell cause an additional error in the kernel,
but this error does not contribute in the Coulomb energy of charge distributions
that are confined to the half-sized Wigner-Seitz cell and are
resolvable on the chosen Fourier grid.\cite{TruncationMIC}

\subsection{Partially-truncated Coulomb kernels}

Next, we generalize the above construction to systems with
lower-dimensional periodicity, where the Coulomb kernel is truncated
along some lattice directions and remains long-ranged along the others.
In these geometries, the kernel is still singular around $\vec{G}=0$,
albeit with a slower divergence: $\ln G$ for one periodic direction
or $1/G$ for two periodic directions in contrast to $1/G^2$ for the fully periodic case.
A general shape for the truncation domain in the non-periodic directions leads to
a kernel which is singular for an entire line or plane of reciprocal
lattice vectors passing through $\vec{G}=0$. Ismail-Beigi\cite{TruncationWS}
pointed out that Wigner-Seitz truncation localizes the singularity
to the single point $\vec{G}=0$ in all these cases.

We can understand the special property of the Wigner-Seitz truncation
by writing the Coulomb kernel truncated on an arbitrary domain $D$ as
\begin{equation}
\tilde{K}\super{D}_{\vec{G}}
 = \int d\vec{r} e^{-i\vec{G}\cdot\vec{r}} \frac{1}{r} \theta_D(\vec{r})
 = \int d\vec{k} \frac{4\pi}{k^2} \tilde{\theta}_D(\vec{G}-\vec{k}),
\label{eqn:TruncatedKernelTransform}
\end{equation}
where $\theta_D(\vec{r})$ is a function that is 1 for $\vec{r}\in D$
and 0 otherwise, and $\tilde{\theta}_D$ is its Fourier transform.
Since $\theta_D(\vec{r})$ is constant along periodic directions,
$\tilde{\theta}$ is zero for wave-vectors with any component
along those directions. In general, the singularity from $4\pi/k^2$
`infects' all wave-vectors with no component along the periodic directions.
Hence, the singularity is spread to a plane of points for one-dimensional or
wire-like systems, and a line of points for two-dimensional or slab-like systems.
However, the Fourier transform of a $\theta$-function with a shape that
tiles with the periodicity of the lattice, such as the Wigner-Seitz cell,
is zero at all non-zero reciprocal lattice vectors.
This confines the singularity to $G=0$ for these special cases.

Now consider, without loss of generality, the slab geometry with its one truncated
direction along $z$, and let $L$ be the unit cell length along that direction.
The Fourier transform of $1/r$ over the two periodic directions evaluates
to $2\pi e^{-G_\rho |z|}/G_\rho$ where $G_\rho$ is the component
of the wave-vector along the untruncated directions.
At $G_\rho=0$, removing the singular part $2\pi/G_\rho$ leaves
behind $-2\pi|z|$, the potential due to an infinite plane of charge
with the arbitrary offset in potential fixed to be zero at the plane.
Note that, although this choice of zero of potential does not
change the total energy for a neutral charge distribution,
care must be exercised to ensure that it be consistent for all
interactions between charged subsystems of an overall neutral system.
Finally, the remaining integral over the Wigner-Seitz
cell in the truncated direction, i.e. $z\in[-L/2,L/2)$,
can also be performed analytically,\cite{TruncationWS,TruncationAnalytic}
so that the Coulomb kernel for truncation in slab geometries becomes
\begin{equation}
\tilde{K}\super{slab}_{\vec{G}} =
	\begin{cases}
		\frac{4\pi}{G^2} \left(1 - \cos\frac{G_z L}{2} \exp\frac{-G_\rho L}{2} \right), & G \neq 0\\
		-\pi L^2/2, & G = 0.
	\end{cases}
\label{eqn:SlabKernel}
\end{equation}

Similarly, for the wire geometry with the single periodic direction along $z$,
the partial Fourier transform of $1/r$ over the periodic direction is $2 K_0(G_z \rho)$,
where $\rho = \sqrt{r^2-z^2}$ is the usual cylindrical coordinate,
and $K_0$ is the modified Bessel function of the second kind.
At $G_z = 0$, removing the logarithmically divergent part leaves behind $-2\ln\rho$
so that the regularized partial Fourier transform of the Coulomb potential at $G_z=k$ is
\begin{equation}
C_k(\rho) \equiv 
\begin{cases}
	2 K_0(k\rho), & k \neq 0 \\
	-2 \ln\rho, & k = 0.
\end{cases}
\label{eqn:C_k}
\end{equation}
The remaining Fourier transform over the two truncated directions is analytically
computable for a cylindrical truncation domain,\cite{TruncationAnalytic}
but that choice spreads the divergence beyond $G=0$ as mentioned previously.
On the other hand, the Fourier transform of (\ref{eqn:C_k}) with a Wigner-Seitz truncation
domain is not known in closed form for any two-dimensional lattice system.\cite{TruncationWS}
Accordingly, we generalize the MIC approach\cite{TruncationMIC} and
approximate the partially-truncated wire-geometry Coulomb kernel by
\begin{multline}
\tilde{K}\super{wire}_{\vec{G}} 
\approx \frac{4\pi}{G^2}\left(1-\exp\frac{-G^2}{4\alpha^2}\right) \\
	+ \frac{\Omega_\perp}{N_{\vec{r}_\perp}} \sum_{\vec{r}_\perp\in\textrm{WS}_\perp}
	e^{-i\vec{G}\cdot\vec{r}_\perp} \bar{C}^{\alpha}_{|G_z|}(r_\perp)
\label{eqn:WireKernel}
\end{multline}
where $r_\perp$ are nodes for the two dimensional Gauss-Fourier quadrature mapped down
to the Wigner Seitz cell, WS$_\perp$, of the truncated directions with area $\Omega_\perp$.

Here, we introduce the smooth, long-ranged special function $\bar{C}^{\alpha}_{k}(\rho)$,
which plays the same role for $C_k(\rho)$ that $\textrm{erf}(\alpha r)/r$
plays for $1/r$ in the fully-truncated case of (\ref{eqn:IsolatedKernel}).
Operationally, this function is defined by the two-dimensional convolution
\begin{flalign}
\bar{C}^{\alpha}_{k}(\rho)
 &\equiv e^{\frac{-k^2}{4\alpha^2}} \left( \frac{\alpha}{\pi}e^{-\alpha^2\rho^2}
	\ast C_k(\rho) \right)\nonumber\\
 &= e^{\frac{-k^2}{4\alpha^2}} \int_0^{\infty} 2\alpha^2\rho'd\rho'
	e^{-\alpha^2(\rho^2+\rho'^2)} I_0(2\alpha^2\rho\rho') C_k(\rho').
\label{eqn:Cbar_k_alpha}
\end{flalign}
For $k = 0$, (\ref{eqn:Cbar_k_alpha}) reduces to the analytical expression
$\bar{C}^{\alpha}_{0}(\rho) = -2\ln\rho-\Gamma_0(\alpha^2\rho^2)$,
but for $k \neq 0$, $\bar{C}^{\alpha}_{k}(\rho)$ needs to be
parametrized numerically.\footnote{Efficient subroutines for evaluating
$\bar{C}^{\alpha}_{k}(\rho)$ and constructing truncated kernels are available
as a part of the open source density-functional software, JDFTx.\cite{JDFTx}}
The choice of $\alpha$ and FFT resolution in (\ref{eqn:Cbar_k_alpha})
follow the discussion for the fully-truncated MIC construction,
except that $R\sub{in}$ is the radius of the two dimensional
Wigner-Seitz cell WS$_\perp$, and independent two-dimensional fast
Fourier transforms produce the results for each plane of constant $G_z$.

\subsection{Ewald sums for reduced-dimensional systems} \label{sec:Ewald}

The plane-wave Coulomb kernels above, truncated over the Wigner-Seitz cell in
one, two or three lattice directions, enable the calculation of Coulomb
interaction energies in slab, wire and isolated geometries respectively.
However, a purely reciprocal-space method is only practical if at least
one of the two charge densities, $\rho_1(\vec{r})$ or $\rho_2(\vec{r})$
in (\ref{eqn:CoulombEnergyPW}), is bandwidth limited.
The interaction energy of point nuclei with each other does not satisfy this
criterion and requires the use of an Ewald sum.\cite{Ewald3D}
Generalizing the standard Ewald method to an arbitrary combination of
truncated and periodic lattice directions, gives the interaction energy for a
set of point charges $Z_i$ at locations $\vec{r}_i$ in the first unit cell, as
\begin{multline}
E\sub{ewald} = \sum_{\substack{\vec{R},i,j \\ i \neq j \textrm{ if } \vec{R}=0}} \frac{Z_i Z_j}{2}
	\frac{\textrm{erfc }\eta |\vec{r}_i+\vec{R}-\vec{r}_j|}{|\vec{r}_i+\vec{R}-\vec{r}_j|} \\
+ \sum_{\vec{G},i,j} \frac{Z_i Z_j}{2\Omega\sub{per}}
	e^{-i\vec{G}\cdot(\vec{r}_i-\vec{r}_j)} g^\eta_{\vec{G}}(\vec{r}_i-\vec{r}_j)
- \frac{\eta}{\sqrt{\pi}}\sum_i Z_i^2.
\label{eqn:Ewald}
\end{multline}
Here, the first term evaluates the contribution due to the short-ranged part
$\textrm{erfc}(\eta r)/r$ of the Coulomb potential, the second term captures
the contribution due to the remaining long-ranged part $\textrm{erf}(\eta r)/r$,
and the third term exactly cancels the self interactions introduced by the second term.
The standard range-separation parameter $\eta$ is adjusted to simultaneously optimize
the convergence of the sum over lattice vectors $\vec{R}$ as well as that over
reciprocal lattice vectors $\vec{G}$. (When some lattice directions are truncated,
$\vec{R}$ and $\vec{G}$ correspond to the lattice vectors and reciprocal lattice
vectors of the lower dimensional Bravais lattice of periodic directions alone.)
Finally, in the second term, $\Omega\sub{per}$ is the volume, area or length of
the unit cell along the periodic directions alone and $g^\eta_{\vec{G}}(\vec{r})$
is related to the (partial) Fourier transform over those directions
of the long-ranged part of the Coulomb potential.

When all three directions are periodic, $g^\eta_{\vec{G}}(\vec{r}) = \exp\frac{-G^2}{4\eta^2}$,
the double sum over point charges factorizes to the square of the structure factor,
and (\ref{eqn:Ewald}) reduces to the standard Ewald sum.\cite{Ewald3D}
Next, for the slab geometry truncated, without loss of generality, along the z direction, we find
\begin{equation}
g^\eta_{\vec{G}}(\vec{r}) =
\begin{cases}
	\frac{\pi}{G}
	\left( f_G^\eta(z) + f_G^\eta(-z) \right),
& G\neq 0 \\
	- 2\pi\left(z~\textrm{erf}(\eta z) + \frac{e^{-\eta^2 z^2}}{\eta\sqrt{\pi}}\right),
& G = 0,
\end{cases}
\label{eqn:EwaldSlab}
\end{equation}
where $f_G^\eta(z) \equiv e^{Gz} \textrm{erfc} \left( G/2\eta + \eta z \right)$,
and this reduces (\ref{eqn:Ewald}) to the `Ewald 2D' formula.\cite{Ewald2D,Ewald2Dtest}

The Ewald sum for the wire geometry with one periodic and two truncated directions
does not seem to have been addressed previously, perhaps because
$g^\eta_{\vec{G}}$ is not analytically expressible in that case.
In fact, we can show that $g^\eta_{\vec{G}}(\vec{r}) = \bar{C}^\eta_G(\sqrt{r^2-z^2})$,
precisely the function defined in (\ref{eqn:Cbar_k_alpha}), which was
introduced for our generalization of the MIC method to this geometry.
Finally, when all three lattice directions are truncated, the Coulomb
kernel has no $G=0$ singularity, and an Ewald sum is not required.
In this case, the Coulomb energy of a set of point charges is computed
directly in real space as a sum over all pairs in one unit cell.


\begin{thebibliography}{10}%
\makeatletter
\providecommand \@ifxundefined [1]{%
 \ifx #1\undefined \expandafter \@firstoftwo
 \else \expandafter \@secondoftwo
\fi
}%
\providecommand \@ifnum [1]{%
 \ifnum #1\expandafter \@firstoftwo
 \else \expandafter \@secondoftwo
\fi
}%
\providecommand \enquote [1]{``#1''}%
\providecommand \bibnamefont  [1]{#1}%
\providecommand \bibfnamefont [1]{#1}%
\providecommand \citenamefont [1]{#1}%
\providecommand\href[0]{\@sanitize\@href}%
\providecommand\@href[1]{\endgroup\@@startlink{#1}\endgroup\@@href}%
\providecommand\@@href[1]{#1\@@endlink}%
\providecommand \@sanitize [0]{\begingroup\catcode`\&12\catcode`\#12\relax}%
\@ifxundefined \pdfoutput {\@firstoftwo}{%
 \@ifnum{\z@=\pdfoutput}{\@firstoftwo}{\@secondoftwo}%
}{%
 \providecommand\@@startlink[1]{\leavevmode}%
 \providecommand\@@endlink[0]{}%
}{%
 \providecommand\@@startlink[1]{%
  \leavevmode
  \pdfstartlink
   attr{/Border[0 0 1 ]/H/I/C[0 1 1]}%
   user{/Subtype/Link/A<</Type/Action/S/URI/URI(#1)>>}%
  \relax
 }%
 \providecommand\@@endlink[0]{\pdfendlink}%
}%
\providecommand \url  [0]{\begingroup\@sanitize \@url }%
\providecommand \@url [1]{\endgroup\@href {#1}{\urlprefix}}%
\providecommand \urlprefix [0]{URL }%
\providecommand \Eprint[0]{\href }%
\@ifxundefined \urlstyle {%
  \providecommand \doi [1]{doi:\discretionary{}{}{}#1}%
}{%
  \providecommand \doi [0]{doi:\discretionary{}{}{}\begingroup
  \urlstyle{rm}\Url }%
}%
\providecommand \doibase [0]{http://dx.doi.org/}%
\providecommand \Doi[1]{\href{\doibase#1}}%
\providecommand \bibAnnote [3]{%
  \BibitemShut{#1}%
  \begin{quotation}\noindent
    \textsc{Key:}\ #2\\\textsc{Annotation:}\ #3%
  \end{quotation}%
}%
\providecommand \bibAnnoteFile [2]{%
  \IfFileExists{#2}{\bibAnnote {#1} {#2} {\input{#2}}}{}%
}%
\providecommand \typeout [0]{\immediate \write \m@ne }%
\providecommand \selectlanguage [0]{\@gobble}%
\providecommand \bibinfo [0]{\@secondoftwo}%
\providecommand \bibfield [0]{\@secondoftwo}%
\providecommand \translation [1]{[#1]}%
\providecommand \BibitemOpen[0]{}%
\providecommand \bibitemStop [0]{}%
\providecommand \bibitemNoStop [0]{.\EOS\space}%
\providecommand \EOS [0]{\spacefactor3000\relax}%
\providecommand \BibitemShut [1]{\csname bibitem#1\endcsname}%
\bibitem{HK-DFT}%
  \BibitemOpen
  \bibfield{author}{%
  \bibinfo {author} {\bibfnamefont{P.}~\bibnamefont{Hohenberg}}\ and\ \bibinfo
  {author} {\bibfnamefont{W.}~\bibnamefont{Kohn}},\ }%
  \bibfield{journal}{%
  \bibinfo {journal} {Phys. Rev.}\ }%
  \textbf{\bibinfo {volume} {136}},\ \bibinfo {pages} {B864} (\bibinfo {year}
  {1964})%
  \bibAnnoteFile{NoStop}{HK-DFT}%
\bibitem{KS-DFT}%
  \BibitemOpen
  \bibfield{author}{%
  \bibinfo {author} {\bibfnamefont{W.}~\bibnamefont{Kohn}}\ and\ \bibinfo
  {author} {\bibfnamefont{L.}~\bibnamefont{Sham}},\ }%
  \bibfield{journal}{%
  \bibinfo {journal} {Phys. Rev.}\ }%
  \textbf{\bibinfo {volume} {140}},\ \bibinfo {pages} {A1133} (\bibinfo {year}
  {1965})%
  \bibAnnoteFile{NoStop}{KS-DFT}%
\bibitem{GW}%
  \BibitemOpen
  \bibfield{author}{%
  \bibinfo {author} {\bibfnamefont{M.~S.}\ \bibnamefont{Hybertsen}}\ and\
  \bibinfo {author} {\bibfnamefont{S.~G.}\ \bibnamefont{Louie}},\ }%
  \bibfield{journal}{%
  \bibinfo {journal} {Phys. Rev. B}\ }%
  \textbf{\bibinfo {volume} {34}},\ \bibinfo {pages} {5390} (\bibinfo {year}
  {1986})%
  \bibAnnoteFile{NoStop}{GW}%
\bibitem{BSE}%
  \BibitemOpen
  \bibfield{author}{%
  \bibinfo {author} {\bibfnamefont{E.~E.}\ \bibnamefont{Salpeter}}\ and\
  \bibinfo {author} {\bibfnamefont{H.~A.}\ \bibnamefont{Bethe}},\ }%
  \bibfield{journal}{%
  \bibinfo {journal} {Phys. Rev.}\ }%
  \textbf{\bibinfo {volume} {84}},\ \bibinfo {pages} {1232} (\bibinfo {year}
  {1951})%
  \bibAnnoteFile{NoStop}{BSE}%
\bibitem{DFTinaccuracy}%
  \BibitemOpen
  \bibfield{author}{%
  \bibinfo {author} {\bibfnamefont{X.}~\bibnamefont{Xu}}\ and\ \bibinfo
  {author} {\bibfnamefont{W.~A.}\ \bibnamefont{Goddard}},\ }%
  \bibfield{journal}{%
  \bibinfo {journal} {J. Chem. Phys.}\ }%
  \textbf{\bibinfo {volume} {121}},\ \bibinfo {pages} {4068} (\bibinfo {year}
  {2004})%
  \bibAnnoteFile{NoStop}{DFTinaccuracy}%
\bibitem{B3LYP}%
  \BibitemOpen
  \bibfield{author}{%
  \bibinfo {author} {\bibfnamefont{A.~D.}\ \bibnamefont{Becke}},\ }%
  \bibfield{journal}{%
  \bibinfo {journal} {J. Chem. Phys.}\ }%
  \textbf{\bibinfo {volume} {98}},\ \bibinfo {pages} {5648} (\bibinfo {year}
  {1993})%
  \bibAnnoteFile{NoStop}{B3LYP}%
\bibitem{PBE0}%
  \BibitemOpen
  \bibfield{author}{%
  \bibinfo {author} {\bibfnamefont{C.}~\bibnamefont{Adamo}}\ and\ \bibinfo
  {author} {\bibfnamefont{V.}~\bibnamefont{Barone}},\ }%
  \bibfield{journal}{%
  \bibinfo {journal} {J. Phys. Chem.}\ }%
  \textbf{\bibinfo {volume} {110}},\ \bibinfo {pages} {6158} (\bibinfo {year}
  {1999})%
  \bibAnnoteFile{NoStop}{PBE0}%
\bibitem{AuxFunc-GygiBaldereschi}%
  \BibitemOpen
  \bibfield{author}{%
  \bibinfo {author} {\bibfnamefont{F.}~\bibnamefont{Gygi}}\ and\ \bibinfo
  {author} {\bibfnamefont{A.}~\bibnamefont{Baldereschi}},\ }%
  \bibfield{journal}{%
  \bibinfo {journal} {Phys. Rev. B}\ }%
  \textbf{\bibinfo {volume} {34}},\ \bibinfo {pages} {4405} (\bibinfo {year}
  {1986})%
  \bibAnnoteFile{NoStop}{AuxFunc-GygiBaldereschi}%
\bibitem{AuxFunc-Wenzien}%
  \BibitemOpen
  \bibfield{author}{%
  \bibinfo {author} {\bibfnamefont{B.}~\bibnamefont{Wenzien}}, \bibinfo
  {author} {\bibfnamefont{G.}~\bibnamefont{Cappellini}},\ and\ \bibinfo
  {author} {\bibfnamefont{F.}~\bibnamefont{Bechstedt}},\ }%
  \bibfield{journal}{%
  \bibinfo {journal} {Phys. Rev. B}\ }%
  \textbf{\bibinfo {volume} {51}},\ \bibinfo {pages} {14701} (\bibinfo {year}
  {1995})%
  \bibAnnoteFile{NoStop}{AuxFunc-Wenzien}%
\bibitem{AuxFunc-Carrier}%
  \BibitemOpen
  \bibfield{author}{%
  \bibinfo {author} {\bibfnamefont{P.}~\bibnamefont{Carrier}}, \bibinfo
  {author} {\bibfnamefont{S.}~\bibnamefont{Rohra}},\ and\ \bibinfo {author}
  {\bibfnamefont{A.}~\bibnamefont{G\"orling}},\ }%
  \bibfield{journal}{%
  \bibinfo {journal} {Phys. Rev. B}\ }%
  \textbf{\bibinfo {volume} {75}},\ \bibinfo {pages} {205126} (\bibinfo {year}
  {2007})%
  \bibAnnoteFile{NoStop}{AuxFunc-Carrier}%
\bibitem{HSE03}%
  \BibitemOpen
  \bibfield{author}{%
  \bibinfo {author} {\bibfnamefont{J.}~\bibnamefont{Heyd}}, \bibinfo {author}
  {\bibfnamefont{G.~E.}\ \bibnamefont{Scuseria}},\ and\ \bibinfo {author}
  {\bibfnamefont{M.}~\bibnamefont{Ernzerhof}},\ }%
  \bibfield{journal}{%
  \bibinfo {journal} {J. Chem. Phys.}\ }%
  \textbf{\bibinfo {volume} {118}},\ \bibinfo {pages} {8207} (\bibinfo {year}
  {2003})%
  \bibAnnoteFile{NoStop}{HSE03}%
\bibitem{HSE12}%
  \BibitemOpen
  \bibfield{author}{%
  \bibinfo {author} {\bibfnamefont{J.~E.}\ \bibnamefont{Moussa}}, \bibinfo
  {author} {\bibfnamefont{P.~A.}\ \bibnamefont{Schultz}},\ and\ \bibinfo
  {author} {\bibfnamefont{J.~R.}\ \bibnamefont{Chelikowsky}},\ }%
  \bibfield{journal}{%
  \bibinfo {journal} {J. Chem. Phys.}\ }%
  \textbf{\bibinfo {volume} {136}},\ \bibinfo {pages} {204117} (\bibinfo {year}
  {2012})%
  \bibAnnoteFile{NoStop}{HSE12}%
\bibitem{SphericalTruncation}%
  \BibitemOpen
  \bibfield{author}{%
  \bibinfo {author} {\bibfnamefont{J.}~\bibnamefont{Spencer}}\ and\ \bibinfo
  {author} {\bibfnamefont{A.}~\bibnamefont{Alavi}},\ }%
  \bibfield{journal}{%
  \bibinfo {journal} {Phys. Rev. B}\ }%
  \textbf{\bibinfo {volume} {77}},\ \bibinfo {pages} {193110} (\bibinfo {year}
  {2008})%
  \bibAnnoteFile{NoStop}{SphericalTruncation}%
\bibitem{QMC-MPC}%
  \BibitemOpen
  \bibfield{author}{%
  \bibinfo {author} {\bibfnamefont{A.~J.}\ \bibnamefont{Williamson}}, \bibinfo
  {author} {\bibfnamefont{G.}~\bibnamefont{Rajagopal}}, \bibinfo {author}
  {\bibfnamefont{R.~J.}\ \bibnamefont{Needs}}, \bibinfo {author}
  {\bibfnamefont{L.~M.}\ \bibnamefont{Fraser}}, \bibinfo {author}
  {\bibfnamefont{W.~M.~C.}\ \bibnamefont{Foulkes}}, \bibinfo {author}
  {\bibfnamefont{Y.}~\bibnamefont{Wang}},\ and\ \bibinfo {author}
  {\bibfnamefont{M.~Y.}\ \bibnamefont{Chou}},\ }%
  \bibfield{journal}{%
  \bibinfo {journal} {Phys. Rev. B}\ }%
  \textbf{\bibinfo {volume} {55}},\ \bibinfo {pages} {R4851} (\bibinfo {year}
  {1997})%
  \bibAnnoteFile{NoStop}{QMC-MPC}%
\bibitem{ModelAnalysisEXX}%
  \BibitemOpen
  \bibfield{author}{%
  \bibinfo {author} {\bibfnamefont{N.~A.~W.}\ \bibnamefont{Holzwarth}}\ and\
  \bibinfo {author} {\bibfnamefont{X.}~\bibnamefont{Xu}},\ }%
  \bibfield{journal}{%
  \bibinfo {journal} {Phys. Rev. B}\ }%
  \textbf{\bibinfo {volume} {84}},\ \bibinfo {pages} {113102} (\bibinfo {year}
  {2011})%
  \bibAnnoteFile{NoStop}{ModelAnalysisEXX}%
\bibitem{MonkhorstPack}%
  \BibitemOpen
  \bibfield{author}{%
  \bibinfo {author} {\bibfnamefont{H.~J.}\ \bibnamefont{Monkhorst}}\ and\
  \bibinfo {author} {\bibfnamefont{J.~D.}\ \bibnamefont{Pack}},\ }%
  \bibfield{journal}{%
  \bibinfo {journal} {Phys. Rev. B}\ }%
  \textbf{\bibinfo {volume} {13}},\ \bibinfo {pages} {5188} (\bibinfo {year}
  {1976})%
  \bibAnnoteFile{NoStop}{MonkhorstPack}%
\bibitem{Note1}%
  \BibitemOpen
  \bibinfo {note} {For a uniform $k$-point mesh, the difference mesh is uniform
  and $\Gamma $-centered even if the original mesh is off-$\Gamma $.}%
  \bibAnnoteFile{Stop}{Note1}%
\bibitem{AuxFunc-DucheminGygi}%
  \BibitemOpen
  \bibfield{author}{%
  \bibinfo {author} {\bibfnamefont{I.}~\bibnamefont{Duchemin}}\ and\ \bibinfo
  {author} {\bibfnamefont{F.}~\bibnamefont{Gygi}},\ }%
  \bibfield{journal}{%
  \bibinfo {journal} {Comp. Phys. Comm}\ }%
  \textbf{\bibinfo {volume} {181}},\ \bibinfo {pages} {855} (\bibinfo {year}
  {2010})%
  \bibAnnoteFile{NoStop}{AuxFunc-DucheminGygi}%
\bibitem{WannierEXX}%
  \BibitemOpen
  \bibfield{author}{%
  \bibinfo {author} {\bibfnamefont{X.}~\bibnamefont{Wu}}, \bibinfo {author}
  {\bibfnamefont{A.}~\bibnamefont{Selloni}},\ and\ \bibinfo {author}
  {\bibfnamefont{R.}~\bibnamefont{Car}},\ }%
  \bibfield{journal}{%
  \bibinfo {journal} {Phys. Rev. B}\ }%
  \textbf{\bibinfo {volume} {79}},\ \bibinfo {pages} {085102} (\bibinfo {year}
  {2009})%
  \bibAnnoteFile{NoStop}{WannierEXX}%
\bibitem{MLWF}%
  \BibitemOpen
  \bibfield{author}{%
  \bibinfo {author} {\bibfnamefont{N.}~\bibnamefont{Marzari}}\ and\ \bibinfo
  {author} {\bibfnamefont{D.}~\bibnamefont{Vanderbilt}},\ }%
  \bibfield{journal}{%
  \bibinfo {journal} {Phys. Rev. B}\ }%
  \textbf{\bibinfo {volume} {56}},\ \bibinfo {pages} {12847} (\bibinfo {year}
  {1997})%
  \bibAnnoteFile{NoStop}{MLWF}%
\bibitem{WannierLocalization}%
  \BibitemOpen
  \bibfield{author}{%
  \bibinfo {author} {\bibfnamefont{W.}~\bibnamefont{Kohn}},\ }%
  \bibfield{journal}{%
  \bibinfo {journal} {Phys. Rev.}\ }%
  \textbf{\bibinfo {volume} {115}},\ \bibinfo {pages} {809} (\bibinfo {year}
  {1959})%
  \bibAnnoteFile{NoStop}{WannierLocalization}%
\bibitem{DensityMatrixLocalization}%
  \BibitemOpen
  \bibfield{author}{%
  \bibinfo {author} {\bibfnamefont{S.}~\bibnamefont{Ismail-Beigi}}\ and\
  \bibinfo {author} {\bibfnamefont{T.~A.}\ \bibnamefont{Arias}},\ }%
  \bibfield{journal}{%
  \bibinfo {journal} {Phys. Rev. Lett.}\ }%
  \textbf{\bibinfo {volume} {82}},\ \bibinfo {pages} {2127} (\bibinfo {year}
  {1999})%
  \bibAnnoteFile{NoStop}{DensityMatrixLocalization}%
\bibitem{TruncationMIC}%
  \BibitemOpen
  \bibfield{author}{%
  \bibinfo {author} {\bibfnamefont{G.~J.}\ \bibnamefont{Martyna}}\ and\
  \bibinfo {author} {\bibfnamefont{M.~E.}\ \bibnamefont{Tuckerman}},\ }%
  \bibfield{journal}{%
  \bibinfo {journal} {J. Chem. Phys}\ }%
  \textbf{\bibinfo {volume} {110}},\ \bibinfo {pages} {2810} (\bibinfo {year}
  {1999})%
  \bibAnnoteFile{NoStop}{TruncationMIC}%
\bibitem{ProbeChargeEwald}%
  \BibitemOpen
  \bibfield{author}{%
  \bibinfo {author} {\bibfnamefont{J.}~\bibnamefont{Paier}}, \bibinfo {author}
  {\bibfnamefont{R.}~\bibnamefont{Hirschl}}, \bibinfo {author}
  {\bibfnamefont{M.}~\bibnamefont{Marsman}},\ and\ \bibinfo {author}
  {\bibfnamefont{G.}~\bibnamefont{Kresse}},\ }%
  \bibfield{journal}{%
  \bibinfo {journal} {J. Chem. Phys.}\ }%
  \textbf{\bibinfo {volume} {122}},\ \bibinfo {pages} {234102} (\bibinfo {year}
  {2005})%
  \bibAnnoteFile{NoStop}{ProbeChargeEwald}%
\bibitem{Ewald3D}%
  \BibitemOpen
  \bibfield{author}{%
  \bibinfo {author} {\bibfnamefont{P.}~\bibnamefont{Ewald}},\ }%
  \bibfield{journal}{%
  \bibinfo {journal} {Ann. Phys.}\ }%
  \textbf{\bibinfo {volume} {369}},\ \bibinfo {pages} {253} (\bibinfo {year}
  {1921})%
  \bibAnnoteFile{NoStop}{Ewald3D}%
\bibitem{HSE06}%
  \BibitemOpen
  \bibfield{author}{%
  \bibinfo {author} {\bibfnamefont{A.~V.}\ \bibnamefont{Krukau}}, \bibinfo
  {author} {\bibfnamefont{O.~A.}\ \bibnamefont{Vydrov}}, \bibinfo {author}
  {\bibfnamefont{A.~F.}\ \bibnamefont{Izmaylov}},\ and\ \bibinfo {author}
  {\bibfnamefont{G.~E.}\ \bibnamefont{Scuseria}},\ }%
  \bibfield{journal}{%
  \bibinfo {journal} {J. Chem. Phys}\ }%
  \textbf{\bibinfo {volume} {125}},\ \bibinfo {pages} {224106} (\bibinfo {year}
  {2006})%
  \bibAnnoteFile{NoStop}{HSE06}%
\bibitem{HSEconvergence}%
  \BibitemOpen
  \bibfield{author}{%
  \bibinfo {author} {\bibfnamefont{J.}~\bibnamefont{Paier}}, \bibinfo {author}
  {\bibfnamefont{M.}~\bibnamefont{Marsman}}, \bibinfo {author}
  {\bibfnamefont{K.}~\bibnamefont{Hummer}}, \bibinfo {author}
  {\bibfnamefont{G.}~\bibnamefont{Kresse}}, \bibinfo {author}
  {\bibfnamefont{I.~C.}\ \bibnamefont{Gerber}},\ and\ \bibinfo {author}
  {\bibfnamefont{J.~G.}\ \bibnamefont{Angyan}},\ }%
  \bibfield{journal}{%
  \bibinfo {journal} {J. Chem. Phys.}\ }%
  \textbf{\bibinfo {volume} {124}} (\bibinfo {year} {154709})%
  \bibAnnoteFile{NoStop}{HSEconvergence}%
\bibitem{PBE}%
  \BibitemOpen
  \bibfield{author}{%
  \bibinfo {author} {\bibfnamefont{J.~P.}\ \bibnamefont{Perdew}}, \bibinfo
  {author} {\bibfnamefont{K.}~\bibnamefont{Burke}},\ and\ \bibinfo {author}
  {\bibfnamefont{M.}~\bibnamefont{Ernzerhof}},\ }%
  \bibfield{journal}{%
  \bibinfo {journal} {Phys. Rev. Lett.}\ }%
  \textbf{\bibinfo {volume} {77}},\ \bibinfo {pages} {3865} (\bibinfo {year}
  {1996})%
  \bibAnnoteFile{NoStop}{PBE}%
\bibitem{JDFTx}%
  \BibitemOpen
  \bibfield{author}{%
  \bibinfo {author} {\bibfnamefont{R.}~\bibnamefont{Sundararaman}}, \bibinfo
  {author} {\bibfnamefont{K.}~\bibnamefont{Letchworth-Weaver}},\ and\ \bibinfo
  {author} {\bibfnamefont{T.~A.}\ \bibnamefont{Arias}},\ }%
  \enquote{\bibinfo {title} {{JDFTx}},}\ \bibinfo {howpublished}
  {\url{http://jdftx.sourceforge.net}} (\bibinfo {year} {2012})%
  \bibAnnoteFile{NoStop}{JDFTx}%
\bibitem{Lattice-AlphaSiC}%
  \BibitemOpen
  \bibfield{author}{%
  \bibinfo {author} {\bibfnamefont{R.~F.}\ \bibnamefont{Adamski}}\ and\
  \bibinfo {author} {\bibfnamefont{K.~M.}\ \bibnamefont{Merz}},\ }%
  \bibfield{journal}{%
  \bibinfo {journal} {Z. Kristallogr.}\ }%
  \textbf{\bibinfo {volume} {111}},\ \bibinfo {pages} {350} (\bibinfo {year}
  {1959})%
  \bibAnnoteFile{NoStop}{Lattice-AlphaSiC}%
\bibitem{Lattice-BetaSiC}%
  \BibitemOpen
  \bibfield{author}{%
  \bibinfo {author} {\bibfnamefont{A.}~\bibnamefont{Taylor}}\ and\ \bibinfo
  {author} {\bibfnamefont{R.~M.}\ \bibnamefont{Jones}},\ }%
  \emph{\bibinfo {title} {Silicon Carbide - A High Temperature Semiconductor}}\
  (\bibinfo {publisher} {Pergamon Press},\ \bibinfo {year} {1960})\ p.\
  \bibinfo {pages} {147}%
  \bibAnnoteFile{NoStop}{Lattice-BetaSiC}%
\bibitem{IceGroundState}%
  \BibitemOpen
  \bibfield{author}{%
  \bibinfo {author} {\bibfnamefont{Z.}~\bibnamefont{Raza}}, \bibinfo {author}
  {\bibfnamefont{D.}~\bibnamefont{Alfe}}, \bibinfo {author}
  {\bibfnamefont{C.~G.}\ \bibnamefont{Salzmann}}, \bibinfo {author}
  {\bibfnamefont{J.}~\bibnamefont{Klime\v{s}}}, \bibinfo {author}
  {\bibfnamefont{A.}~\bibnamefont{Michaelidesade}},\ and\ \bibinfo {author}
  {\bibfnamefont{B.}~\bibnamefont{Slater}},\ }%
  \bibfield{journal}{%
  \bibinfo {journal} {Phys. Chem. Chem. Phys.}\ }%
  \textbf{\bibinfo {volume} {13}},\ \bibinfo {pages} {19788} (\bibinfo {year}
  {2011})%
  \bibAnnoteFile{NoStop}{IceGroundState}%
\bibitem{CRC-Handbook}%
  \BibitemOpen
  \emph{\bibinfo {title} {CRC Handbook of Physics and Chemistry 93\super{rd}
  ed}},\ edited by\ \bibinfo {editor} {\bibfnamefont{W.~M.}\
  \bibnamefont{Haynes}}\ (\bibinfo {year} {2012})\ pp.\ \bibinfo {pages}
  {12:15--12:18}%
  \bibAnnoteFile{NoStop}{CRC-Handbook}%
\bibitem{DFT-review}%
  \BibitemOpen
  \bibfield{author}{%
  \bibinfo {author} {\bibfnamefont{M.~C.}\ \bibnamefont{Payne}}, \bibinfo
  {author} {\bibfnamefont{M.~P.}\ \bibnamefont{Teter}}, \bibinfo {author}
  {\bibfnamefont{D.~C.}\ \bibnamefont{Allan}}, \bibinfo {author}
  {\bibfnamefont{T.~A.}\ \bibnamefont{Arias}},\ and\ \bibinfo {author}
  {\bibfnamefont{J.~D.}\ \bibnamefont{Joannopoulos}},\ }%
  \bibfield{journal}{%
  \bibinfo {journal} {Rev. Mod. Phys.}\ }%
  \textbf{\bibinfo {volume} {64}},\ \bibinfo {pages} {1045} (\bibinfo {year}
  {1992})%
  \bibAnnoteFile{NoStop}{DFT-review}%
\bibitem{TruncationAnalytic}%
  \BibitemOpen
  \bibfield{author}{%
  \bibinfo {author} {\bibfnamefont{C.~A.}\ \bibnamefont{Rozzi}}, \bibinfo
  {author} {\bibfnamefont{D.}~\bibnamefont{Varsano}}, \bibinfo {author}
  {\bibfnamefont{A.}~\bibnamefont{Marini}}, \bibinfo {author}
  {\bibfnamefont{E.~K.~U.}\ \bibnamefont{Gross}},\ and\ \bibinfo {author}
  {\bibfnamefont{A.}~\bibnamefont{Rubio}},\ }%
  \bibfield{journal}{%
  \bibinfo {journal} {Phys. Rev. B},\ \bibinfo {pages} {205119}}%
   (\bibinfo {year} {2006})%
  \bibAnnoteFile{NoStop}{TruncationAnalytic}%
\bibitem{TruncationWS}%
  \BibitemOpen
  \bibfield{author}{%
  \bibinfo {author} {\bibfnamefont{S.}~\bibnamefont{Ismail-Beigi}},\ }%
  \bibfield{journal}{%
  \bibinfo {journal} {Phys. Rev. B}\ }%
  \textbf{\bibinfo {volume} {73}},\ \bibinfo {pages} {233103} (\bibinfo {year}
  {2006})%
  \bibAnnoteFile{NoStop}{TruncationWS}%
\bibitem{Note2}%
  \BibitemOpen
  \bibinfo {note} {Efficient subroutines for evaluating $\protect \mathaccentV
  {bar}016{C}^{\alpha }_{k}(\rho )$ and constructing truncated kernels are
  available as a part of the open source density-functional software,
  JDFTx.\cite {JDFTx}}%
  \bibAnnoteFile{NoStop}{Note2}%
\bibitem{Ewald2D}%
  \BibitemOpen
  \bibfield{author}{%
  \bibinfo {author} {\bibfnamefont{D.~M.}\ \bibnamefont{Heyes}}, \bibinfo
  {author} {\bibfnamefont{M.}~\bibnamefont{Barber}},\ and\ \bibinfo {author}
  {\bibfnamefont{J.~H.~R.}\ \bibnamefont{Clarke}},\ }%
  \bibfield{journal}{%
  \bibinfo {journal} {J. Chem. Soc. Faraday Trans. II}\ }%
  \textbf{\bibinfo {volume} {73}},\ \bibinfo {pages} {1485} (\bibinfo {year}
  {1977})%
  \bibAnnoteFile{NoStop}{Ewald2D}%
\bibitem{Ewald2Dtest}%
  \BibitemOpen
  \bibfield{author}{%
  \bibinfo {author} {\bibfnamefont{E.}~\bibnamefont{Spohr}},\ }%
  \bibfield{journal}{%
  \bibinfo {journal} {J. Chem. Phys}\ }%
  \textbf{\bibinfo {volume} {107}},\ \bibinfo {pages} {6342} (\bibinfo {year}
  {1997})%
  \bibAnnoteFile{NoStop}{Ewald2Dtest}%
\end{thebibliography}

%

\end{document}